\documentclass[pdflatex,sn-nature]{sn-jnl}

\usepackage{graphicx}%
\usepackage{multirow}%
\usepackage{amsmath,amssymb,amsfonts}%
\usepackage{amsthm}%
\usepackage{mathrsfs}%
\usepackage[title]{appendix}%
\usepackage{xcolor}%
\usepackage{textcomp}%
\usepackage{manyfoot}%
\usepackage{booktabs}%
\usepackage{algorithm}%
\usepackage{algorithmicx}%
\usepackage{algpseudocode}%
\usepackage{listings}%
\usepackage{chemformula}%

\theoremstyle{thmstyleone}%
\theoremstyle{thmstyletwo}%

\theoremstyle{thmstylethree}%

\begin{document}

\title[Detection of a four-carbon sugar in interstellar space]{Detection of a four-carbon sugar in interstellar space}


\author*[1]{\fnm{Izaskun} \sur{Jim\'enez-Serra}}\email{ijimenez@cab.inta-csic.es}

\author[2]{\fnm{Juan} \sur{Garc\'{\i}a de la Concepci\'on}}
\author[3]{\fnm{Herma M.} \sur{Cuppen}}
\author[1,4]{\fnm{Marta} \sur{Rey-Montejo}}
\author[1,5]{\fnm{Miguel} \sur{Sanz-Novo}}
\author[1]{\fnm{V\'{\i}ctor M.} \sur{Rivilla}}
\author[1]{\fnm{Jes\'us} \sur{Mart\'{\i}n-Pintado}}
\author[1]{\fnm{Andr\'es} \sur{Meg\'{\i}as}}
\author[1]{\fnm{Carlos} \sur{Briones}}
\author[1,4]{\fnm{David} \sur{San Andr\'es}}
\author[1]{\fnm{Laura} \sur{Colzi}}
\author[6]{\fnm{Shaoshan} \sur{Zeng}}
\author[7,8]{\fnm{Sergio} \sur{Mart\'{\i}n}}
\author[3]{\fnm{Joseph} \sur{Salaris}}
\author[1,6]{\fnm{Antonio} \sur{Mart\'{\i}nez-Henares}}
\author[1]{\fnm{\'Alvaro} \sur{L\'opez-Gallifa}}
\author[9]{\fnm{Miguel} \sur{Requena-Torres}}
\author[10,11]{\fnm{Bel\'en} \sur{Tercero}}
\author[11]{\fnm{Pablo} \sur{de Vicente}}
\author[12,13]{\fnm{Aran} \sur{Insausti}}
\author[14]{\fnm{Elena R.} \sur{Alonso}}
\author[12,13]{\fnm{Emilio J.} \sur{Cocinero}}

\affil*[1]{\orgdiv{Centro de Astrobiolog\'{\i}a (CAB)}, \orgname{CSIC-INTA}, \orgaddress{\street{Carretera de Ajalvir km 4}, \city{Torrej\'on de Ardoz}, \postcode{E--28850}, \state{Madrid}, \country{Spain}}}
\affil[2]{Departamento de Qu\'{\i}mica Org\'anica e Inorg\'anica, Facultad de Ciencias and IACYS-Green Chemistry and Sustainable Development Unit, University of Extremadura, E--06006, Badajoz, Spain}
\affil[3]{Institute for Molecules and Materials, Radboud University, Heyendaalseweg 135, 6525 AJ, Nijmegen, The Netherlands}
\affil[4]{Departamento de F\'{\i}sica de la Tierra y Astrof\'{\i}sica, Facultad de Ciencias F\'{\i}sicas, Universidad Complutense de Madrid, E--28040, Madrid, Spain}
\affil[5]{Center for Astrochemical Studies, Max-Planck-Institut f\"{u}r extraterrestrische Physik,
Giessenbachstrasse 1, Garching bei Munchen, 85748, Germany}
\affil[6]{Star and Planet Formation Laboratory, Cluster for Pioneering Research, RIKEN, 2-1 Hirosawa, Wako, Saitama, 351-0198, Japan}
\affil[7]{European Southern Observatory, Alonso de Córdova 3107, Vitacura 763 0355, Santiago, Chile}
\affil[8]{Joint ALMA Observatory, Alonso de Córdova 3107, Vitacura 763 0355, Santiago, Chile}
\affil[9]{Department of Physics, Astronomy and Geosciences, Towson University, Towson, MD 21252, USA}
\affil[10]{Observatorio Astron\'omico Nacional (OAN-IGN), Calle Alfonso XII, 3, 28014 Madrid, Spain}
\affil[11]{Observatorio de Yebes (OY-IGN), Cerro de la Palera SN, Yebes, Guadalajara, E-19141, Spain}
\affil[12]{Departamento de Química Física, Facultad de Ciencia y Tecnología, Universidad del País Vasco (UPV/EHU), E--48080, Bilbao, Spain}
\affil[13]{Instituto Biofisika (CSIC, UPV/EHU), E--48080, Bilbao, Spain}
\affil[14]{Grupo de Espectroscop\'{\i}a Molecular (GEM), Edificio Quifima, Laboratorios de Espectroscop\'{\i}a y Bioespectroscop\'{\i}a, Universidad de Valladolid, E--47011, Valladolid, Spain}




\abstract{Sugars are essential biomolecules, serving as metabolic fuels, nucleic acid backbone components, and structural or energy-storage polymers. A central question in origin-of-life research is how monosaccharides formed on the primitive Earth, as laboratory experiments under prebiotic conditions yield insufficient concentrations. The detection of ribose, glucose and other monosaccharides in asteroids and meteorites suggests an exogenous origin, possibly in the interstellar medium (ISM) prior to meteoritic parent-body formation. However, no sugar has been observed in the ISM so far. We report the discovery of erythrulose, a chiral four-carbon ketose, in the ISM. The detection has been achieved thanks to ultrasensitive, broadband spectral surveys toward the Galactic Center molecular cloud G+0.693-0.027 obtained using the Yebes 40m and IRAM 30m telescopes. Erythrulose appears to be at least eight times more abundant than analogous three-carbon sugars, which remain undetected in our ultrasensitive observations. Quantum chemical and astrochemical models indicate that erythrulose forms efficiently on interstellar dust grains from simpler two-carbon aldehydes and alcohols. As ketoses readily isomerize into aldoses in aqueous conditions, interstellar erythrulose could have contributed to the sugar inventory available for early metabolic and replication processes.}




\maketitle

\section{Introduction}\label{sec:intro}

The interstellar medium (ISM) is an impressive chemical factory where more than 340 molecules have been detected so far \cite{mcguire2022}, including large aromatic species \cite{cernicharo2021,wenzel2025,fuentetaja2026}. Several of these species, such as urea, hydroxylamine or ethanolamine \cite{belloche2019,jimenez2020,Rivilla2020,rivilla2021a, Rivilla2022a}, are directly connected to origin-of-life chemistry because they are considered precursors of ribonucleosides and lipids \cite{ruiz-mirazo2014,becker2019,menor-salvan2020,fiore2022}. Sugars are also central molecules to prebiotic chemistry \cite{powner2009,patel2015,becker2019}; however, they are usually introduced ad hoc (i.e., as inputs) in prebiotic reaction schemes because their formation under plausible early-Earth conditions remains inefficient
\cite{mizuno1974,kopetzki2011,delidovich2014,haas2020}.

The detection of bio-essential sugars such as ribose and glucose in primitive meteorites and in asteroid Bennu, suggests that at least part of the sugar inventory available on the early Earth may have had an exogenous origin\cite{furukawa2019,Furukawa2026}. A natural possibility is that these sugars, or their precursors, formed before parent-body accretion. However, no sugar has so far been reported in the ISM. Glycolaldehyde (\ch{HOCH2CHO}) is widespread in interstellar space \cite{hollis2000,beltran2009,jorgensen2012} and has often been discussed in this context because of its structural relationship to aldose sugars. Yet, it is a hydroxyaldehyde rather than a true saccharide  \cite{Porterfield2016}. 
 
A major obstacle to searching for interstellar sugars has been the lack of accurate gas-phase rotational data. Such data are essential for astronomical identification, but obtaining them was long considered extremely challenging because sugars are thermally fragile and strongly hygroscopic, which hampers their manipulation and conventional vaporization. This limitation has recently been overcome using ultrafast laser vaporization, which enabled the gas-phase rotational characterization of ribose, 2-deoxyribose, and erythrulose\cite{cocinero2012,pena2013,insausti2021}. These laboratory data opened the way to sensitive  astronomical searches for sugars with three and four carbon atoms (i.e. C3 and C4 sugars) -- e.g., glyceraldehyde (\ch{HOCH2CH(OH)CHO}), dihydroxyacetone (\ch{HOCH2COCH2OH}), 
and erythrulose (\ch{HOCH2CH(OH)COCH2OH}) --, although previous searches remained unsuccessful \cite{hollis2004,widicus-weaver2005,apponi2006,jimenez2020,insausti2021}.

Here, we report the discovery in the ISM of erythrulose, the only C4 ketose, toward the molecular cloud G+0.693-0.027, hereafter G+0.693, located in the Galactic Center region at a distance of $\sim$8.2 kpc from us \cite{gravity2019}. The identification was enabled by an ultrasensitive broadband spectral survey with the Yebes 40m and the IRAM 30m radiotelescopes covering more than 91 GHz across the 7 mm, 3 mm, and 2 mm atmospheric windows (see Methods). G+0.693 is one of the richest molecular reservoirs known in the Galaxy \cite{requena2006,Zeng2018,jimenez2020}, and has yielded numerous detections of new interstellar species of prebiotic interest in recent years \cite[][]{rivilla2021a,Jimenez2022,sanznovo2023}.



\section{Results}\label{sec:results}

\subsection{Detection of erythrulose in G+0.693}\label{sec:detection}

\begin{figure}[ht]
\centering
\includegraphics[width=1.0\textwidth]{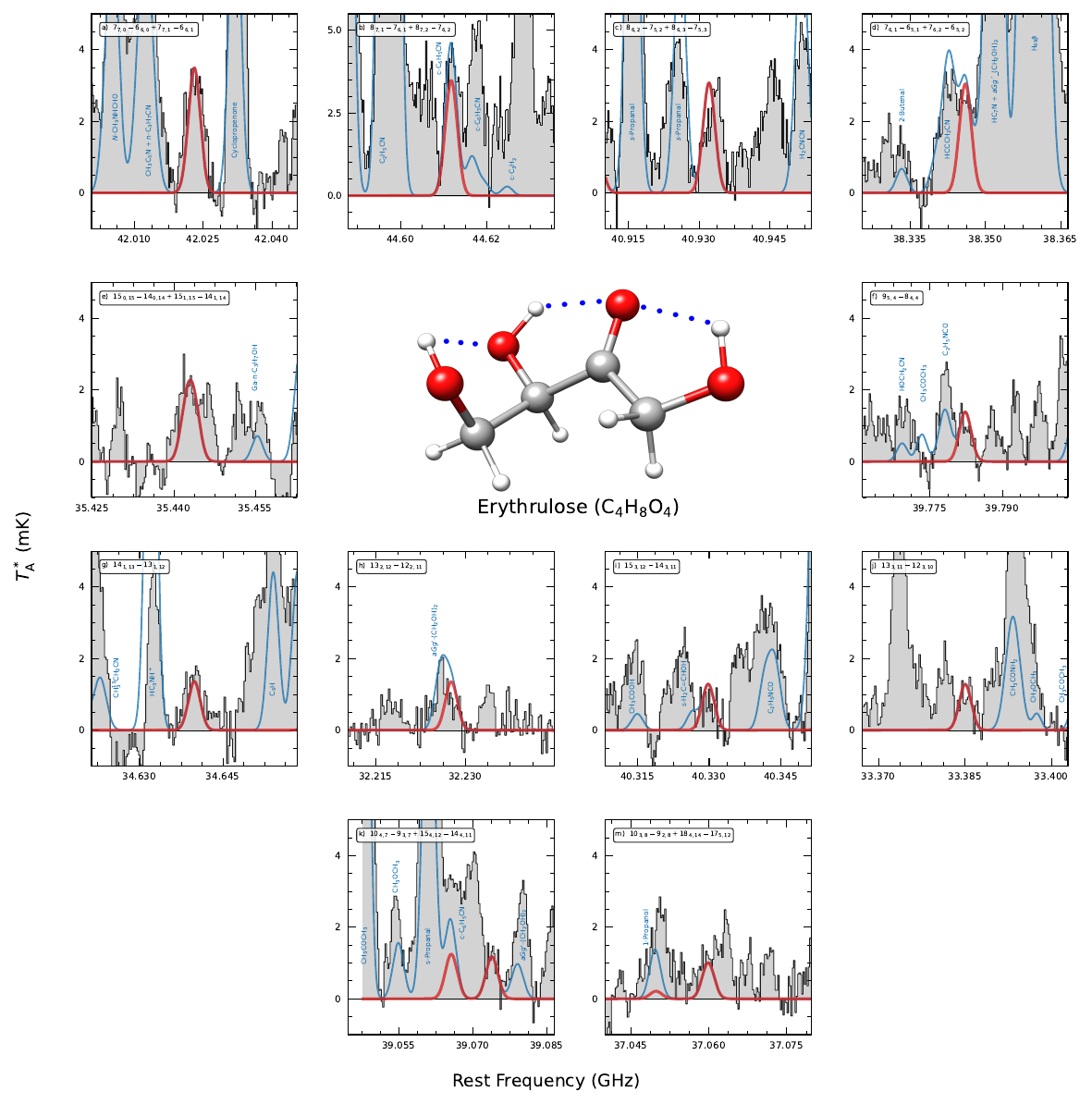}
\caption{{\bf{Brightest and most unblended transitions of erythrulose observed toward the G+0.693 molecular cloud.}} Filled histograms report the observed spectra, red lines show the line profiles of the erythrulose transitions fitted with \textsc{Madcuba}-SLIM, and blue lines present the total fit to the spectra considering all the molecules identified toward the cloud. The quantum numbers of each transition of erythrulose are given in the upper part of each panel. In blue, we also label the molecular species contributing to the observed spectra in the vicinity of the erythrulose lines. The transitions are sorted from the brightest to the weakest lines according to the LTE model.}\label{fig:ery}
\end{figure}

Although line blending remains a challenge in chemically rich sources such as G+0.693, the analysis presented here identifies a set of transitions that are consistent with the predicted spectrum of erythrulose.
Figure \ref{fig:ery} shows 12 sets of lines of erythrulose (accounting for a total of 17 individual transitions) identified in the G+0.693 cloud using the \href{https://cab.inta-csic.es/madcuba/}{\textsc{Madcuba}-SLIM software} (see also Table \ref{tab:spec} in the Extended Data). These correspond to the brightest and most unblended transitions of erythrulose present in our dataset with integrated intensities $\geq$9$\sigma$ -- with $\sigma$ = rms $\times$ $\sqrt{\delta v \times \mathrm{FWHM}}$, where rms is the RMS noise level of the spectra, $\delta$$v$ is the velocity resolution, and FWHM the full width half-maximum of the lines --, which are used to fit the erythrulose emission under the assumption of local thermodynamic equilibrium (LTE) using \textsc{Madcuba}-SLIM (Methods). The individual contribution of the erythrulose transitions is shown in red, while in blue we report the total spectrum predicted by \textsc{Madcuba}-SLIM considering the contribution from all molecular species identified and modelled toward G+0.693 (more than 180 species, including isotopologues, are considered in the model). Note that transitions from higher vibrational states are not expected in this cloud because the molecular emission is sub-thermally excited with measured $T_{\rm ex}$$\leq$15 K\cite{Zeng2018,zeng2020}. 

Six sets of transitions out of the 12 shown in Figure \ref{fig:ery} are classified as predominantly unblended (at 42023 MHz, 40932 MHz, 35442 MHz, 39782 MHz, 40329 MHz, and 33385 MHz), with contamination levels $\leq$25\% (Table \ref{tab:spec}; the criteria to establish the predominantly unblended features of erythrulose are described in Methods and Supplementary Information).   
This ensures that the emission in these features is dominated by erythrulose and provides a robust basis for identification. The transitions at 34639 MHz, 39073 MHz, and 37060 MHz are classified as blended with U-lines according to our selection criteria (Methods and Table \ref{tab:spec}). However, a visual inspection shows that the erythrulose fit predicted by \textsc{Madcuba}-SLIM reproduces well the observed spectra (Figure \ref{fig:ery}). The features at 44611 MHz, 38346 MHz and 32227 MHz are blended with known molecular species but the agreement with the global fit is excellent, further reinforcing the detection of erythrulose. 

Figure \ref{fig:additional-lines} shows the remaining transitions of erythrulose present in our dataset with peak intensities $\geq$1.2 mK, i.e. 3 $\times$ the average rms noise level in the spectra shown in Figure \ref{fig:ery}. These transitions are weaker or more affected by blending and/or noise and are hence not used for the LTE fit of erythrulose. Despite not being used for the LTE fit, all these erythrulose lines predicted by \textsc{Madcuba}-SLIM are consistent with the observed spectra. Given the physical properties of erythrulose in G+0.693 (see below), its brightest lines are expected to fall within the frequency range covered by Yebes 40m (31-50 GHz), as observed for other COMs\cite{Araki2026}. However, some transitions also appear at 3$\,$mm (blue asterisks in Figure \ref{fig:additional-lines}), although they are largely blended. 
The agreement across the full set of detected transitions provides strong, independent validation of the identification.

The fit of the 12 sets of erythrulose transitions shown in Figure \ref{fig:ery} using the \textsc{Madcuba}-SLIM tool yields an excitation temperature of $T_{\rm ex} = 11.3 \pm 1.8$ K and a column density of $N = (8.7 \pm 0.8) \times 10^{13}$ cm$^{-2}$ for a fixed central radial velocity $v_{\rm LSR}$ = 69 km s$^{-1}$ and a linewidth FWHM = 22 km s$^{-1}$ (Methods). 
The derived column density of $(8.7 \pm 0.8)\times10^{13}$ cm$^{-2}$  translates into an erythrulose abundance of $(6.4 \pm 0.6)\times10^{-10}$ assuming a \ch{H2} column density of 1.35$\times$10$^{23}$ cm$^{-2}$ for G+0.693 \cite{martin2008}. 
Glycolaldehyde (with 2 C atoms) is observed with an abundance in G+0.693 similar to that of erythrulose (Table \ref{tab:abun}). 
In contrast, the only C3 sugars, glyceraldehyde and dihydroxyacetone, are not detected in this cloud, with upper limits to their abundance $\leq 4\times10^{-11}$ and $\leq 7\times10^{-11}$, respectively. Erythrulose appears to be $\geq$8-17 times more abundant than C3 sugars (Table \ref{tab:abun}). This is striking, since other chemical families (e.g., alcohols, thiols, aldehydes, isocyanates) show an abundance decrease of roughly one order of magnitude with each added carbon atom\cite[][]{Rodriguez2021a,Rodriguez2021b,Jimenez2022,sanz-novo2022}. One might consider the triol compound glycerol (\ch{HOCH2CH(OH)CH2OH}) as a potential precursor of erythrulose, but this molecule was not detected in our survey either (Table \ref{tab:abun}).

We can estimate the confidence level of the erythrulose detection in G+0.693 following refs.\cite{snyder2005,halfen2006}. Assuming the confusion limit and Gaussian line profiles, the chance of finding a feature at the $v_{\rm LSR}$ of G+0.693 with a tolerance of $\pm$4 km s$^{-1}$ (i.e. the largest velocity shift observed for the molecular line emission in G+0.693)\cite{Jimenez2022,colzi2022} is $p$ = 36\% for a velocity coverage equivalent to the FWHM of the erythrulose lines. The detection of each successive line has a probability $p^n$ with $n$ the number of unblended transitions. The probability of chance alignment for the six most unblended transitions is then $p^6$ = 0.2\%, providing strong statistical support for the identification of erythrulose. It is important to stress that, although G+0.693 presents line-rich and complex spectra, the molecular emission in this cloud is sub-thermally excited with low $T_{\rm ex}$, which yields much lower levels of line blending and line confusion as compared to hotter sources such as massive hot cores and low-mass hot corinos. Even if three or four unblended lines were considered in our analysis, we would still obtain confidence levels of 95.2\% and 98.3\%, respectively. Note that this estimate is conservative since line confusion has not been reached in our G+0.693 spectral surveys.

\subsection{Formation mechanism on interstellar ice surfaces}\label{sec:quantum}

The C3 sugars glyceraldehyde and dihydroxyacetone are factors $\geq$8-17 less abundant than erythrulose, which suggests that they are unlikely to be its dominant precursors under the physical conditions of G+0.693. Therefore, we turn our attention to smaller and more abundant building blocks of this C4 sugar such as glycolaldehyde and ethylene glycol. Recent laboratory experiments of irradiated CH$_3$OH ices reveal that sugars with up to 6 carbon atoms can be produced from smaller sugar and sugar-derivative fragments (i.e. C6 sugars would form from C3 sugar precursors) \cite{zhang2024}. By inspecting the chemical structure of erythrulose, one can envision its formation via the combination of two C2 fragments: 
 \ch{CH2OHC^.O} and \ch{C^.HOHCH2OH}. 
 The first fragment is the glycolaldehyde radical (g*), and the second is the ethylene glycol radical (e*), which contains a pro-chiral center at carbon \ch{C^.}, responsible for the chirality of erythrulose. Both glycolaldehyde and ethylene glycol are present in G+0.693 at high abundances -- their measured abundances are $(6.9 \pm 0.2)\times10^{-10}$ and $(1.7 \pm 0.4)\times10^{-9}$, 1.1 and 2.7 times the erythrulose abundance, respectively (Table \ref{tab:abun}). Moreover, the radical g* has been identified as the main intermediate species formed in laboratory experiments simulating the reaction of atomic H with glycolaldehyde in solid para-\ch{H2} (dense cloud conditions) \cite{joshi2024}. Besides, sugar acids have also been synthesized experimentally by the energetic processing of interstellar ice analogues containing ethylene glycol and CO$_2$, through the formation of the radical e* \cite{Wang2024}. Therefore, the combination of g* and e* could lead to the synthesis of erythrulose in the ISM.

\begin{figure}[ht]
\centering
\includegraphics[width=1.0\textwidth]{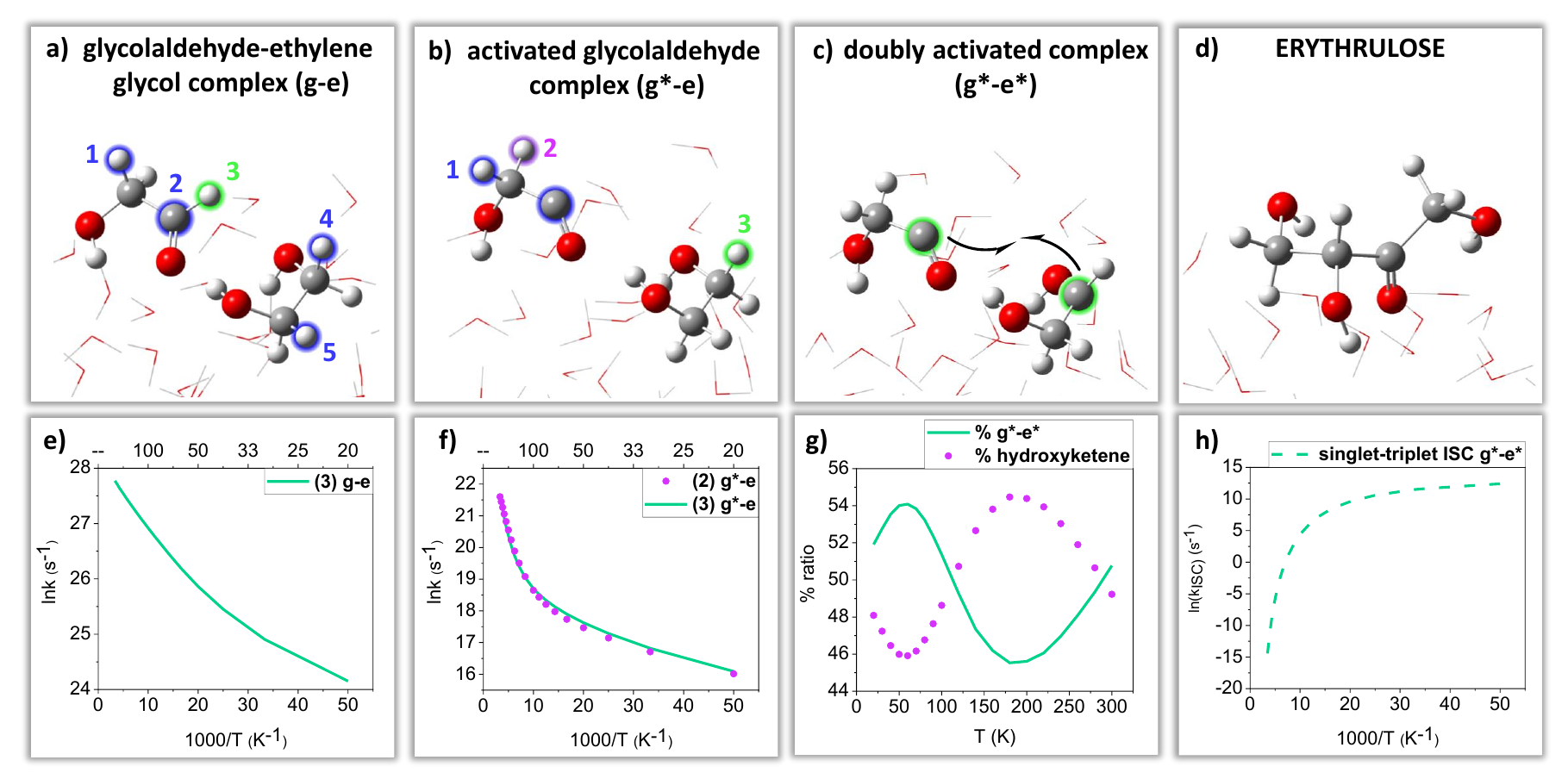}
\caption{{\bf Reaction mechanism of erythrulose from glycolaldehyde (g) and ethylene glycol (e) on ASW.} Optimized geometries are shown for the complexes g-e (panel {\bf a}), the first activated g*-e complex (panel {\bf b}), the doubly activated g*-e* complex (panel {\bf c}), and erythrulose (panel {\bf d}). The available hydrogen and carbon atoms for abstraction, addition and recombination reactions are highlighted in blue, green and purple. The green and purple highlighted hydrogen atoms in panels {\bf a} and {\bf b} represent, respectively, the most viable reactions. Green carbon atoms in panel {\bf c} represent the recombination of the two radicals. Panels {\bf e} and {\bf f} show the Arrhenius plots of the thermal rate constants for reaction 3-g-e (panel {\bf a}) and reactions 2-g*-e and 3-g*-e (panel {\bf b}), respectively. Panel {\bf g} shows the branching ratios for reactions 2-g*-e and 3-g*-e leading to the complexes hydroxyketene-ethylene glycol (not shown) and g*-e*. The rate constant for the ISC of the g*-e* complex is shown in panel {\bf h}. Some water molecules have been removed from the images of the optimized geometries for the sake of clarity.}\label{fig:quantum}
\end{figure}

Figure \ref{fig:quantum} illustrates the formation mechanism of this C4 ketose sugar on the surface of amorphous solid water ice (ASW) from the radicals g* and e*. This formation mechanism has two isoenergetic mirror symmetry mechanisms that form the two enantiomers with 50\% probability each. Panel {\bf a} shows the most favorable orientation of the glycolaldehyde-ethylene glycol complex (g-e) on the ASW and all possible reactive collisions of atomic H with the complex. Among these, the hydrogen abstraction reaction 3-g-e, is the fastest reaction (shown in green, panel {\bf a}), with an energy barrier of 4.1 kJ mol$^{-1}$ and a reaction energy of 27.4 kJ mol$^{-1}$ (Zero-point Energy - $ZPE$ - corrected; see the Supplementary Information for further details). 
The quantum tunneling-corrected rate constant of reaction 3-g-e ranges from 3.1$\times$10$^{10}$ s$^{-1}$ to 1.2$\times$10$^{12}$ s$^{-1}$ in the temperature range between 20--300 K (panel {\bf e} in Figure \ref{fig:quantum}), and hence it is a very fast reaction.

After the H abstraction reaction 3-g-e, the activated glycolaldehyde complex (g*-e, panel {\bf b} of Figure \ref{fig:quantum}) is formed, which could proceed with either subsequent H abstraction reactions or two isomerization reactions by hydrogen migration to the activated carbon of the aldehyde group. The two isomerization reactions present energy barriers of 77.0 and 97.3 kJ mol$^{-1}$, which makes them non-viable under ISM conditions. For the H abstraction reactions, reaction 1-g*-e shows a repulsive potential (see Supplementary Information). However, reactions 2-g*-e and 3-g*-e are energetically favored and form hydroxyketene for reaction 2-g*-e, and the doubly activated complex g*-e* for reaction 3-g*-e (panel {\bf c}). The branching ratios for these two reactions are approximately 50\% in the whole range of temperatures, being slightly higher for the formation of the g*-e* complex at low temperatures (panel {\bf g}). The tunneling-corrected rate constant of reaction 3-g*-e ranges from 9.7$\times$10$^{6}$ s$^{-1}$ to 2.5$\times$10$^{9}$ s$^{-1}$ between 20--300 K.

Once formed, the g*-e* complex is in the triplet state and the radicals cannot recombine since the approach of atoms having parallel electrons induces a repulsive potential. However, the g*-e* complex can experience a spin change under ISM conditions throughout an intersystem crossing process (ISC), since both states are practically degenerate with an energy difference (ZPE corrected) of only 0.04 kJ mol$^{-1}$ in favor of the singlet state (Supplementary Information). The latter process shows rate constants of 2.4$\times$10$^5$ s$^{-1}$--1.6$\times$10$^{-7}$ s$^{-1}$ between 20--300 K, being the slowest process of the whole mechanism (panel {\bf h} in Figure \ref{fig:quantum} and Table \ref{tab:rates}). Finally, the association of the two carbons with antiparallel spin yields erythrulose (panel {\bf d}).

\subsection{Astrochemical modeling of erythrulose formation}\label{sec:modelling}

\begin{figure}[ht]
\centering
\includegraphics[width=1.0\textwidth]{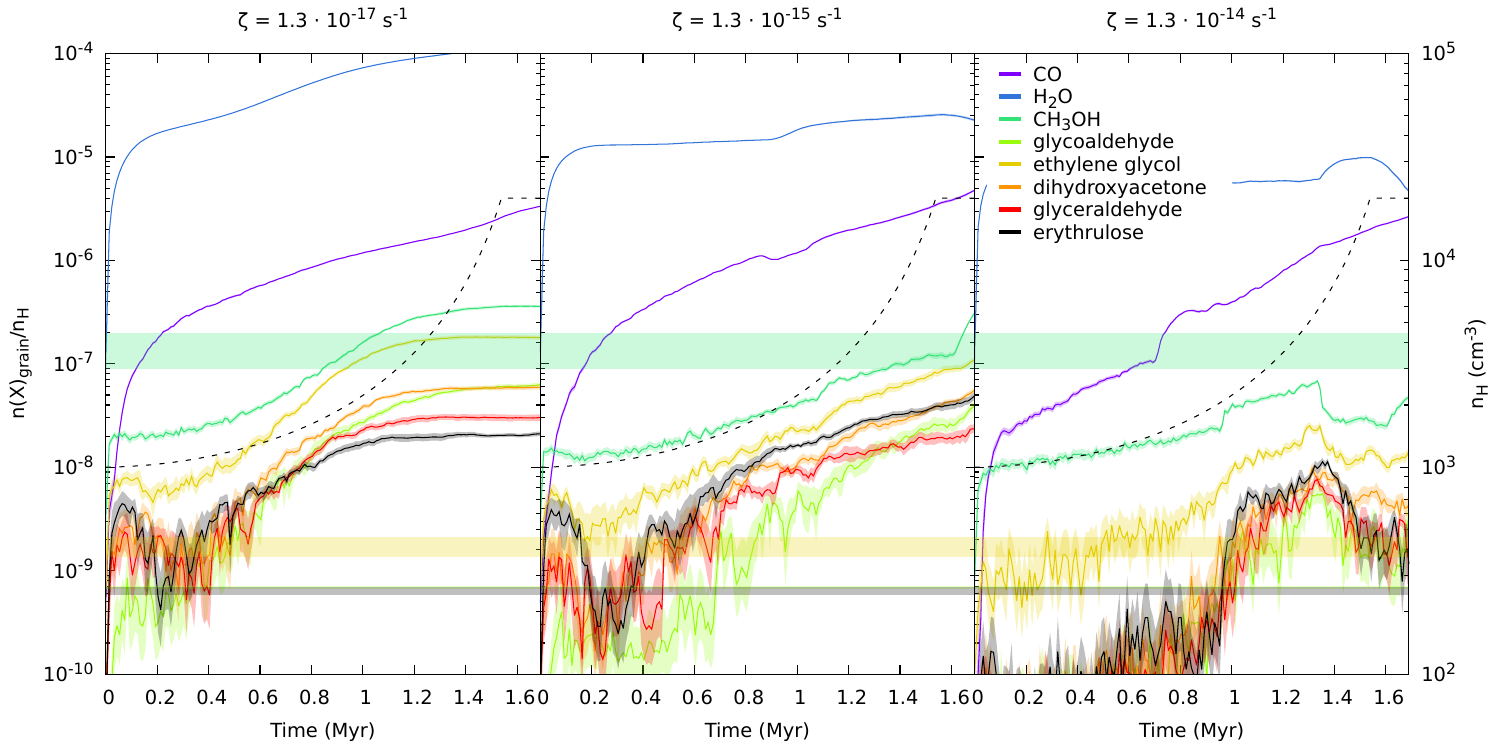}
\caption{{\bf KMC simulations of the ice build-up in a cloud with similar physical conditions to those of G+0.693.} Colour lines indicate the solid ice abundances of CO, H$_2$O, \ch{CH3OH}, glycolaldehyde, ethylene glycol, and all C3 and C4 sugars considered in the model (Methods). Data are presented as mean values $\pm$ 1 standard deviation spread of the nine simulations carried out with different random seeds. Labels are shown in the right panel, and the cosmic-rays ionization rate, $\zeta$, in the upper part of each panel. Dashed line indicates the increase of the gas density during the cloud's collapse. Horizontal color-shaded ranges show the molecular abundance values measured toward G+0.693 together with their 1$\sigma$ uncertainty (Table \ref{tab:abun}).}\label{fig:KMC}
\end{figure}

We have demonstrated that erythrulose can form on the surface of interstellar dust grains from glycolaldehyde and ethylene glycol at a rate constant of (1--2)$\times$10$^{5}$ s$^{-1}$ at the typical dust temperatures measured in Galactic Center molecular clouds (T$_{\rm dust}$$\sim$20-30 K)\cite{rodriguez2004,etxaluze2013}. Using this information, we now evaluate whether we can predict the observed amount of erythrulose in the G+0.693 cloud by implementing the mechanism described above in a Kinetic Monte-Carlo (KMC) simulation code \cite{cuppen2007,cuppen2009,lamberts2014,Simons2020}. We expanded our chemical network to include the formation of other C3 and C4 sugars (glyceraldehyde, dihydroxyacetone, threose, and erythrose) and related compounds (e.g. glycerol). The network allows radicals from C1 and C2 species (e.g., \ch{H^.CO}, \ch{H2C^.OH}, \ch{HC^.OHCHO}, \ch{CHOHC^.O}, and \ch{CH2OHC^.HOH}) to recombine\cite{Simons2020}, forming five different C3 compounds (including the sugars dihydroxyacetone and glyceraldehyde, plus glycerol and two C3 aldehydes) and six different C4 compounds (including threose and erythrulose). Additional hydrogen abstraction, hydrogenation, and UV photodissociation reactions (including those inducing the breaking of the CC bond of glycolaldehyde and ethylene glycol) were added to allow the interconversion between species (see Methods). We ran simulations of the ice build-up for different cosmic-ray ionization rates: $\zeta$ = 1.3$\times$10$^{-17}$ s$^{-1}$ (the standard Galactic value) and 1.3$\times$10$^{-15}$ s$^{-1}$ and 1.3$\times$10$^{-14}$ s$^{-1}$ (100 and 1000 times higher, as applicable to Galactic Center molecular clouds \cite{goto2014,Lepetit2016}). The model assumes an initial hydrogen-nuclei density of $10^3$~$\rm cm^{-3}$, rising to $2\times 10^4$~$\rm cm^{-3}$ over 1.5 Myr during the collapse of the cloud, after which the density is kept constant (Figure~\ref{fig:KMC}). We fix the dust temperature at $T_{\rm dust}$ = 20 K, consistent with observations of G+0.693 \cite{etxaluze2013}. Each simulation was run with nine different random seeds.


Figure \ref{fig:KMC} shows that both C3 and C4 sugars can form efficiently in the ice in our models. Erythrulose is the most efficiently produced C4 sugar, achieving higher abundance than the C4 aldoses (threose and erythrose) under most conditions. Consistent with our observations, erythrulose tends to reach slightly higher abundances than the C3 sugars (glyceraldehyde and dihydroxyacetone), especially at higher $\zeta$ and shorter time scales. This behavior results from the higher UV photo-destruction rate assumed for C3 sugars compared to C4 sugars (Methods). Chemically, smaller sugar molecules are expected to be more easily destroyed than larger ones, since larger molecules have more degrees of freedom to distribute energy upon UV absorption. In our simulations, ethylene glycol is always more abundant than any C3 and C4 sugar, in agreement with our observations. Glycolaldehyde, however, only attains abundances comparable to (or slightly higher than) the C3 and C4 sugars toward the end of the simulations. 

Molecular clouds in the Galactic Center are pervaded by low-velocity shocks from cloud shearing and collisions \cite{martin-pintado1997}. G+0.693 is believed to be undergoing such a collision, driving a large-scale shock of $v_{\rm s}\sim20$ km s$^{-1}$ \cite{zeng2020}. Ice material is thus expected to be partially injected into the gas phase in G+0.693 by grain sputtering \cite{jimenez2008}. Figure \ref{fig:KMC} shows that the KMC models whose ice abundances come closest to the observed gas-phase values in G+0.693 are those with $\zeta\geq$1.3$\times$10$^{-14}$ s$^{-1}$. Such high $\zeta$ were also invoked in previous models of this source to explain the abundances of certain molecular ions (e.g., PO$^+$ and HOCS$^+$) \cite{Rivilla2022b,Sanz2024}. In our simulations, the predicted abundances in the ice of CH$_3$OH, glycolaldehyde, ethylene glycol, and erythrulose, all lie within a factor of 5 of the observed values (Table \ref{tab:abun}) and well within the tolerance of one order of magnitude typically considered in chemical modelling. 
The C3 sugars (glyceraldehyde and dihydroxyacetone) are overproduced in the model by factors $\sim$25-70. However, the abundance ratios of glycolaldehyde, ethylene glycol and of the C3 sugars with respect to erythrulose differ by a bit more than
one order of magnitude (Table \ref{tab:abun2}), being fairly consistent given the large uncertainties in our simulations.

The discrepancy between the modeled abundances in the ice and the observed abundances in the gas could be due to a combination of effects. First, only a fraction of the ices may be sputtered in the shock ($\leq$30\%, based on the KMC results). C3 and C4 sugars are more refractory than water and hence may be harder to eject from grains. Alternatively, C3 and C4 sugars may rapidly re-adsorb onto dust grains given the low dust temperature ($T_{\rm dust}$ = 20 K) in G+0.693. This could result in less of those sugars remaining in the gas to be observed. The detection of CO and CO$_2$ ice in Galactic Center molecular clouds \cite{ginsburg2023} indicates that a fraction of the ices remain on dust grains despite the dust energetic processing. In addition, note that the KMC simulations do not include an extended gas-phase chemical network for these COMs and hence, C3 and C4 sugars may undergo efficient gas-phase destruction through ion-neutral reactions\cite{Sanz2024} after the ice release in the shock. Glyceraldehyde has recently been found to be more unstable than glycolaldehyde in the gas phase, which may explain its lack of detection \cite{lema2025}. Finally, uncertainties in reaction rates (e.g., photodissociation efficiencies)\cite{fernandez2023}, reaction branching ratios, or even unmodeled processes, might also play a role in overestimating sugar abundances in the ice in our model. 

\section{Discussion}\label{sec:discussion}

Erythrulose, with 14 atoms in its structure, represents the largest non-cyclic molecular species identified to date in the ISM \cite{Fried2024}, and the first detected molecule containing four oxygen atoms. 
It is also the first sugar and the second chiral molecule reported in the ISM \cite{mcguire2016}. Its detection not only provides direct evidence that complex, chiral species can form under interstellar conditions, but it also takes us to a higher level in the ladder of interstellar chemical complexity, suggesting that other prebiotic (and potentially chiral) molecules could also form and survive under the extreme conditions of the ISM.

Prebiotic chemistry experiments have shown that ribonucleotides -- the building blocks of RNA -- can be synthesized from mixtures containing sugars such as erythrulose \cite{Islam2017}. Ketoses like erythrulose can readily isomerize into their aldose counterparts (i.e., threose and erythrose) in aqueous environments\cite{nagoski2001}, a transformation that is central to the formation of ribose and then nucleic acids \cite{colville2021}. However, prebiotic experiments have typically introduced such sugars as external inputs. Our detection of erythrulose in the G+0.693 molecular cloud demonstrates that this monosaccharide can be synthesized abiotically under interstellar conditions. 
The similarity of the organic inventories of comets and presolar environments supports a connection between the chemistry observed in the ISM and that inherited by minor bodies of the Solar System\cite{altwegg2019}. Therefore, the detection of interstellar erythrulose suggests that complex sugars may have formed in the proto-solar molecular cloud, being subsequently transferred to minor bodies. 

Sugars such as ribose and glucose have indeed been found in meteorites\cite{furukawa2019} and in asteroid Bennu\cite{Furukawa2026}. In addition, erythrulose has also been proposed to exist on the surface of outer Solar System bodies, including the Kuiper Belt object Arrokoth\cite{zhang2024}. Laboratory irradiation experiments support the interstellar formation of erythrulose, as well as other sugar-derivatives such as sugar acids and sugar alcohols \cite{Wang2024, zhang2024}, in agreement with our astrochemical model. Our astrochemical simulations indeed show that erythrulose can form efficiently in ices under a range of cosmic-ray ionization rates, including those expected in dust traps of protoplanetary disks \cite{ligterink2024}. In these regions, energetic processing transforms simple ices into macromolecular organics, with up to 4\% of the ice reservoir affected\cite{ligterink2024}. Hence, once formed, erythrulose and other sugars may be incorporated into planetesimals (as measured in Arrokoth \cite{zhang2024} and meteorites \cite{furukawa2019}), later contributing to the inventory of prebiotic organic material on planetary surfaces. The discovery of chiral erythrulose also supports the idea that small enantiomeric excesses measured in meteorites (of a few percent)\cite{Cronin1997,Cooper2016} may originate in extraterrestrial environments, which could have contributed to the emergence of biological homochirality on early Earth through subsequent chemical amplification processes \cite{Ribo2019}. 
 
Even under conservative assumptions -- considering only the observed gas-phase abundance of erythrulose in G+0.693 -- 
and given the interstellar water abundance of 10$^{-4}$ (see ref. \cite{whittet1991}), the erythrulose/water abundance ratio measured in G+0.693 is of $\sim$6$\times$10$^{-6}$. Other prebiotic organics such as ethanolamine, also detected in G+0.693, have been found in meteorites with abundance ratios with respect to water similar to those measured in the ISM\cite{rivilla2021a}. Therefore, considering that the average water content in meteorites is 7500 ppm\cite[][]{piani2020} and an estimated amount of organic matter delivered to early Earth of 10$^{16}$-10$^{18}$ kg \cite[][]{chyba1992}, we estimate that $\sim$(0.5-50)$\times$10$^{9}$ kg of erythrulose could have been delivered to early Earth during the Late Heavy Bombardment between 4.1 and 3.9 billion years ago\cite{Koeberl2006}. Although the extent and intensity of the Late Heavy Bombardment has recently been questioned\cite{Cartwright2022}, Monte-Carlo simulations of Earth’s impact history show that biopoiesis -- the process describing the development of living matter from non-living organic matter -- could occur within a much broader timeline between 4.45 and 3.9 billion years ago with at least a 16\% chance\cite{Wogan2024}. Since glycolaldehyde and small sugars and sugar derivatives have been shown to survive meteoritic impacts\cite{mccaffrey2014,zellner2020}, the presence of erythrulose in interstellar space strengthens the case for an exogenous origin of sugars relevant to the synthesis of the first nucleic acids. The presence of erythrulose on the surface of a primitive Earth and its rapid isomerization into threose\cite{nagoski2001} support its involvement in the formation of Threose Nucleic Acid (TNA), a structurally simpler RNA analogue and the simplest of all potential sugar-containing nucleic acids. TNA has been synthesized in the laboratory and, in the context of the origin of life, has been proposed as one of the polymers that could have been involved in a pre-RNA world \cite{Yu2012}. Therefore, the discovery of interstellar erythrulose demonstrates that the ISM could be a viable sugar feedstock for the prebiotic synthesis of the first nucleic acids not only in a primitive Earth but elsewhere in the Universe.

\section*{Methods}

\subsection*{Observational campaign and spectral coverage}

We conducted a broadband, ultrasensitive spectral survey of the Galactic Center molecular cloud G+0.693 using the Yebes 40 m (Guadalajara, Spain) and IRAM 30 m (Granada, Spain) radiotelescopes. Observations were performed in position-switching mode toward $\alpha$(J2000) = 17$^{\rm h}$47$^{\rm m}$2$^{\rm s}$, $\delta$(J2000) = -28$^\circ$21$'$27$''$, with the off position shifted by $\Delta\alpha$ = -885$''$ and $\Delta\delta$ = 290$''$. 
The Yebes 40 m data (project 21A014) were collected between March 2021 to March 2022 using the ultra-broadband Nanocosmos Q-band (7 mm) HEMT receiver, which provides full frequency coverage from 31.07 to 50.42 GHz in two linear polarizations\cite{Tercero2021}. Spectra were recorded using the 16 fast Fourier transform spectrometers (38 kHz channel width), with frequency setups at 41.4 and 42.3 GHz to flag spurious features. Data reduction and averaging were performed using the \href{https://github.com/andresmegias/gildas-class-pipeline/}{Python-based pipeline} developed by ref.\cite{Megías2023} and the \href{https://cab.inta-csic.es/madcuba/index.html}{\textsc{Madcuba} (Madrid Data CUBe Analysis) software} (see ref.\cite{Martin2019}). The final spectra were smoothed to a frequency resolution of 256 kHz (1.5-2.5 km s$^{-1}$). The half power beam width (HPBW) of the telescope ranged between 35” and 55”. All data were calibrated in $T_{\rm A}^*$ units because the molecular emission toward G+0.693 is extended over the beam\cite{requena2006,Zeng2018}. The final rms noise achieved was 0.25-0.9 mK per channel.   

The IRAM 30 m observations (projects 123-22 and 076-23) were performed during different sessions in 2023 (February), and 2024 (March and April) using the Eight Mixer Receivers (EMIR). Spectral coverage spanned 83.2-115.41, 132.28-140.39, and 142-173.81 GHz. Each frequency set up was shifted in frequency in order to identify possible contamination of spurious lines coming from the image band. The fast Fourier Transform Spectrometer (FTS200) provided a spectral resolution of 195 kHz, although the spectra were finally smoothed to 615 kHz (1.0-2.2 km s$^{-1}$). The spectra were calibrated in $T_{\rm A}^*$ units. The HPBW ranged from 14” to 29”. The final rms noise was 0.5-2.5 mK at 3 mm and 1.0-1.6 mK at 2 mm per channel. Frequency gaps were filled using data from previous IRAM surveys\cite{Rodriguez2021a,sanznovo2023}. \\

\subsection*{Identification of Predominantly Unblended Transitions}

The rotational spectrum of the L-1 conformer of erythrulose -- the lowest energy open-chain conformer -- was measured in the laboratory by ref.\cite{insausti2021}. Its spectroscopic entry was calculated using SPCAT up to 720 GHz. Predicted frequencies below 18 GHz have uncertainties as low as 1 kHz, which translates into errors of up to 120 kHz in the Q band, 
equivalent to $\sim$0.7-1.2 km s$^{-1}$. These uncertainties are much smaller than the typical linewidths (of $\sim$20 km s$^{-1}$) observed in G+0.693, and than the spectral resolution of our data (minimum of 256 kHz). A similar approach has been used for the discovery of several other molecular species in this cloud even using spectra at higher frequencies (e.g. at 3mm; see the case of HOCS$^+$)\cite{Sanz2024,lattanzi2024}.

\begin{figure}[htb!]
\centering
\includegraphics[width=\hsize]{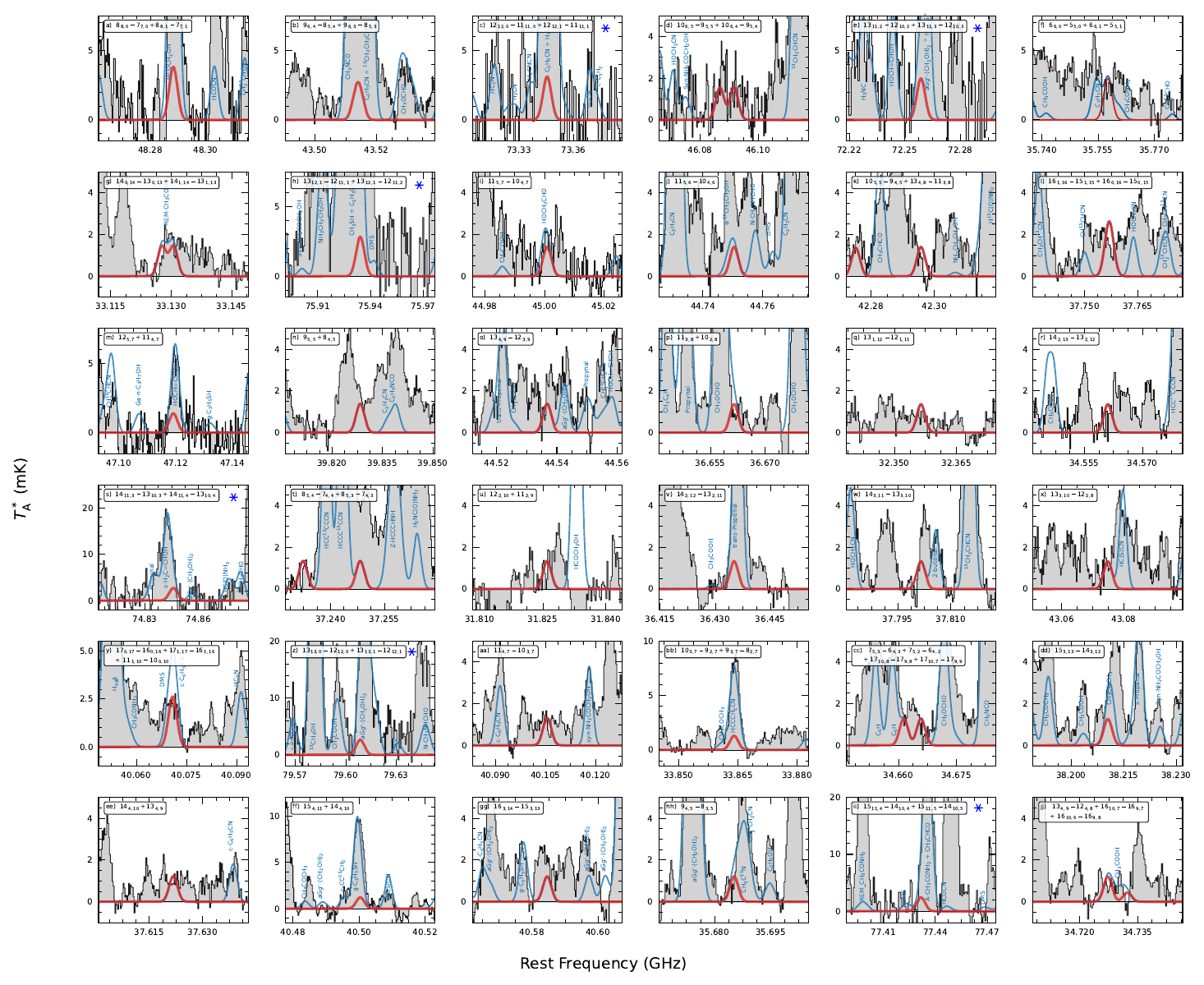}
\caption{\textbf{Remaining erythrulose transitions present in our dataset with peak intensities $\geq$1.2 mK.} Filled histograms report the observed spectra, red lines show the line profiles of the erythrulose transitions fitted with MADCUBA-SLIM, and blue lines present the total fit to the spectra considering all the molecules identified toward G+0.693. The quantum numbers of each transition of erythrulose are given in the upper part of each panel. In blue, we label the molecular species contributing to the observed spectra in the vicinity of the erythrulose lines.}
\label{fig:additional-lines}
\end{figure}

The SPCAT spectroscopy entry was incorporated into the \textsc{Madcuba} package (version 31/05/2024)\cite{Martin2019}, and transitions of erythrulose were identified using the Spectral Line Identification and Modeling (SLIM) tool under the assumption of Local Thermodynamic Equilibrium (LTE). SLIM generates the LTE synthetic spectra for comparison with observed data. 
This allows us to identify the brightest and most unblended transitions of erythrulose shown in Figure \ref{fig:ery}. The line classification has been carried out following the criteria described in ref.\cite{rey-montejo2024}, which are based on the accepted standard of ref.\cite{snyder2005} and ref.\cite{halfen2006}:

i) Accurate rest frequencies: 
The frequency uncertainties of the identified erythrulose transitions are $\leq$120 kHz (see values in parentheses within the first column of Table \ref{tab:spec}), which lies below the spectral resolution of our Yebes 40m and IRAM 30m data with a minimum value of 256 kHz. 

ii) Frequency agreement: $"$An accurate astronomical rest frequency of the assigned transition must be in reasonable agreement with the frequency corresponding to the LSR velocity of the source$"$\cite{snyder2005}. The erythrulose emission is well reproduced by fixing its V$_{\rm LSR}$ to 69 km s$^{-1}$ (Table \ref{tab:abun}), in excellent agreement with the V$_{\rm LSR}$ obtained for other COMs toward G+0.693 \cite{Zeng2018}. 

iii) Linewidth agreement: The derived line width of the erythrulose lines is 22 km s$^{-1}$ (Table \ref{tab:abun}). This value is consistent with those observed for other molecules toward the G+0.693 cloud\cite{Massalkhi2023}.

iv) Beam dilution: 
Our data do not suffer from beam dilution because the emission of COMs such as glycolaldehyde and ethylene glycol toward G+0.693 is extended across the Sgr B2 molecular cloud\cite{Li2017} and therefore, across the single-dish beams of the IRAM 30m and Yebes 40m telescopes. 

v) Relative intensities: 
“Once several molecular transition assignments have been made, their relative intensities must be tested for consistency”\cite{snyder2005}. All observed spectral features are consistent with the LTE predictions obtained with \textsc{Madcuba}-SLIM (Figure \ref{fig:ery} and Figure \ref{fig:additional-lines}).

vi) Confirmation of transitions: “There must be spectral features at all favorable, physically connected transitions over a sufficiently large wavelength range“. Our dataset covers over 91 GHz in the 7mm, 3mm and 2mm wavelength ranges, and all predicted features of erythrulose are consistent with the observed spectra. Note that the global LTE fit overpredicts the spectrum for the 17$_{0,17}$$\rightarrow$16$_{0,16}$ and 17$_{1,17}$$\rightarrow$16$_{1,16}$ transitions at $\sim$40.070 GHz (panel y in Figure \ref{fig:additional-lines}), but this can be considered as an outlier in our model. Other blended transitions such as the 12$_{5,7}$$\rightarrow$11$_{4,7}$ and 15$_{4,11}$$\rightarrow$14$_{4,10}$ transitions (panels m and ff in Figure \ref{fig:additional-lines}), and the 14$_{11,3}$$\rightarrow$13$_{10,3}$+14$_{11,4}$$\rightarrow$13$_{10,4}$ and 10$_{3,7}$$\rightarrow$9$_{2,7}$+9$_{3,7}$$\rightarrow$8$_{2,7}$ line sets (panels s and bb in the same Figure), match very well the observed spectra.

vii) Level of blending: 
To determine the level of blending with other molecules or U-lines, we have obtained the residual area after subtracting the LTE fit  of the erythrulose lines to the total measured area in the observed spectra for the velocity range between V$_{\rm LSR}$ = 69$\pm$FWHM (i.e. over 2$\times$FWHM). This velocity range covers 98.2\% of the total area underneath a Gaussian line profile. As shown in Figure \ref{fig:gaussian}, if one considers the most unfavourable case in which the observed spectral feature is flat due to heavy line blending, the derived residual area would represent $\sim$50\% the total area measured over a velocity range of 2$\times$FWHM. Therefore, we establish a criteria by which an observed feature is considered as mostly unblended if the residual contribution is less than half this limit, or $\leq$25\% (see Supplementary Information). A similar approach has been adopted in recent detections of new molecules in G+0.693, including large complex organics\cite{rey-montejo2024,Araki2026}.

Following these criteria, we identify a total of 6 mostly unblended features, which account for 9 individual transitions of erythrulose. Although blended with residual emission $\geq$25\%, the visual inspection of panels b, d and h in Figure \ref{fig:ery} shows that the global LTE fit reproduces well the observed spectra for the sets of transitions 8$_{7,1}$$\rightarrow$7$_{6,1}$+8$_{7,2}$$\rightarrow$7$_{6,2}$, and 7$_{6,1}$$\rightarrow$6$_{5,1}$+7$_{6,2}$$\rightarrow$6$_{5,2}$, and for the 13$_{2,12}$$\rightarrow$12$_{2,11}$ line. All these transitions, together with the consistency of the rest of lines shown in Figure \ref{fig:ery} and in Figure \ref{fig:additional-lines}, yield a robust assignment of erythrulose. Full line parameters of the transitions shown in Figure \ref{fig:ery} are listed in Table \ref{tab:spec}. \\

\subsection*{LTE Analysis with \textsc{Madcuba}}

We performed a nonlinear least-squares LTE fit of the erythrulose emission by applying the \textsc{Autofit} tool based on the
Levenberg–Marquardt algorithm within SLIM\cite{Martin2019} to derive the best-fit physical parameters: excitation temperature ($T_{\rm ex}$), radial velocity ($v_{\rm LSR}$), line width (FWHM), and column density ($N$). To achieve convergence, the FWHM of the erythrulose emission had to be fixed to 22 km s$^{-1}$, consistent with the typical line profiles in G+0.693\cite{requena2006,Zeng2018}. Derived values are reported in Table 1 and agree with those obtained for other complex organic molecules (COMs) in this source\cite{requena2006,Zeng2018}.


\subsection*{Computational methods}

We simulated the ASW ice by carrying out molecular dynamic simulations of 100 water molecules at 300 K to amorphize the system, which was then cooled at 10 K to preserve the amorphous structure. Since our mechanistic approach involves the non-diffusive association of activated glycolaldehyde (g) and ethylene glycol (e) on ASW ice (see Figure \ref{fig:quantum}), we placed several glycolaldehyde molecules around ethylene glycol and optimized each pair, obtaining only two stable structures: one with a carbonyl–hydroxyl interaction (CO) and another with a hydroxyl–hydroxyl interaction (OH). The two optimized CO and OH structures were used as fragments in a rigid docking study to generate glycolaldehyde–ethylene glycol complexes placed across the ASW surface (Figure \ref{fig:quantum}). Binding energies for both types were similar overall, but the CO complex showed a higher maximum interaction energy, so its strongest-bound structure was selected for the mechanistic study (Figure \ref{fig:Ebind}). Electronic structure calculations were performed with Gaussian 16\cite{gaussian} using a three-layer ONIOM method (PW6B95(D3)/def2-TZVP:PM7R6:UFF)\cite{Zhao2005,pm7,uff,Ahlrichs2005} to model the reaction mechanism on ASW ice. Reactivity was initiated via hydrogen abstraction reactions forming radicals on the surface. The intersystem crossing (ISC) needed to form erythrulose was studied using CASSCF(6e,5o)/def2-TZVP in ORCA\cite{casscf,orca}. The unimolecular rate constants for the hydrogen abstraction reactions were studied using transision state theory (TST) using the Pilgrim software\cite{pilgrim}. Tunneling corrections were applied using the Eckart approximation\cite{eckart1930penetration}. The ISC rate of the g*-e* intermediate was estimated using Marcus' semi-classical theory\cite{Marcus1964,Marcus1985}. The details of these calculations can be found in the Supplementary Information.

\subsection*{Grain-surface chemistry simulations: Kinetic Monte-Carlo (KMC) approach}

We extended a previously established grain-surface network\cite{Simons2020,Ioppolo2021} to include six C3-forming radical-radical reactions (leading to e.g. tartonaldehyde, hydroxypyruvaldehyde, glyceraldehyde, dihydroxyacetone, and glycerol) and seven C4-forming reactions (leading to 2,3-dihydroxybutanedial, 2,4-dihydroxy-3-oxobutanal, tetrose, dimethylolglyxoxal, erythrulose, and tetritol). The rate constants are estimated as $2\times 10^{6}$ s$^{-1}$.
Additional hydrogenation reactions and abstraction reactions ($k = 10^{5}$ s$^{-1}$) were included to allow  interconversion via radical intermediates (ref.\cite{Alvarez-Barcia:2018}). The C3 radicals can react with \ch{HC^.O} and \ch{C^.H2OH} to C4 sugar and sugar-derivative species. The  glycoaldehyde--ethylene glycol complex was also included explicitly in the model to account for the surface reactions outlined in Figure~\ref{fig:quantum} (rate constants from Table \ref{tab:rates}).
All photodissociation processes were modeled using canonical rate constants of the form: \\

\noindent
For C2 species:
\begin{equation}
k = 10^{-9}\exp(-2.5A_V)+500\zeta \qquad\textrm{s}^{-1}
\end{equation}
For the C3 species:
\begin{equation}
k = 9.0\times 10^{-10}\exp(-2.5A_V)+450\zeta \qquad\textrm{s}^{-1}
\end{equation}
For C4 species: 
\begin{equation}
k = 8.1\times 10^{-10}\exp(-2.5A_V)+405\zeta \qquad\textrm{s}^{-1}.
\end{equation}

The slightly reduced photodissociation rates for C4 species reflect their greater resilience due to higher internal degrees of freedom. 
The grain surface chemistry is triggered by deposition of H, \ch{H2}, C, N, O, CO, \ch{N2}, and \ch{O2}. Their fluxes are calculated self-consistently assuming that these are the main elemental gas-phase reservoirs and the remaining is on the grain. \href{https://uclchem.github.io/}{UCLCHEM} (see ref.\cite{holdship2017}) is used to determine the time-dependent gas-phase CO/C, \ch{O2}/O, and \ch{N2}/N ratios and H and \ch{H2} fluxes.

\backmatter



\bmhead{Code Availability}

The \textsc{Madcuba} software is publicly available at https://cab.inta-csic.es/madcuba/. A description of the package is provided in ref.\cite{Martin2019}. The python-based script developed by ref. \cite{Megías2023} is available at https://github.com/andresmegias/gildas-class-python/.
UCLCHEM can be downloaded freely from https://uclchem.github.io/. The KMC code is available on reasonable request to H. Cuppen. 



\bmhead{Acknowledgements}

We thank the Yebes 40m and IRAM 30m staff for their support during the observations. Based on observations carried out with the Yebes 40 m telescope (project 21A014) and the IRAM 30m telescope (projects 123-22, 076-23). The 40 m radio telescope at Yebes Observatory is operated by the Spanish Geographic Institute (IGN; Ministerio de Transportes y Movilidad Sostenible). IRAM is supported by INSU/CNRS (France), MPG (Germany) and IGN (Spain). Computational resources were provided by CENITS and Foundation Computaex through the High-Performance Computing facility LUSITANIA-II, which are greatly appreciated.

\bmhead{Author contributions} 
I.J.-S. initiated and managed the project. She wrote the observational part of the manuscript, together with the abstract, introduction and concluding parts. J.G.C. wrote the part of the quantum chemical calculations, while H.C. carried out the KMC simulations and wrote that part of the manuscript. M.R.-M. contributed creating the figures and tables and provided comments to an initial draft. M.S-N., V.M.R., J.M.-P., L.C., S.Z., S.M., M.R.-T., B.T., P.d.V., A.M.-H., A.M., A. L.-G. and D.S.A. contributed to the acquisition and reduction of the astronomical data as well as providing comments to the manuscript. C.B. investigated the biochemical relevance of erythrulose and other sugars in the context of the origin of life and contributed to the manuscript. E.J.C., A.I., and E.R.A. provided the SPCAT-format spectroscopic predictions of erythrulose, enabling its identification in the astronomical data. E.J.C. thoroughly revised a previous version of the manuscript, and together with A.I. and E.R.A. also commented on the paper.  

\bmhead{Competing Interests}
The authors declare no competing interests. 

\bmhead{Funding}
This work is supported by ERC grant OPENS, GA No. 101125858 funded by the European Union. Views and opinions expressed are however those of the author(s) only and do not necessarily reflect those of the European Union or the European Research Council Executive Agency. Neither the European Union nor the granting authority can be held responsible for them. I.J-.S., J.G.dlC., M.S.-N., V.M.R., J.M.-P., L.C., S.Z., S.M., A.M., A.M.-H., A.L.-G., M.R.-T., D.S.A also acknowledge partial support from grant number PID2022-136814NB-I00 funded by the Spanish Ministry of Science, Innovation and Universities/State Agency of Research MICIU/AEI/ 10.13039/501100011033 and by $"$ERDF/EU$"$. I.J-.S., A.M. and S.Z. acknowledge partial support from the CSIC Bilateral project SOULMATE (BIJSP25017). 
J.G.dlC. acknowledges support from European Funds for Regional Development and the Autonomous Government of Extremadura (Grant No. GR24020).
M.S.-N. also acknowledges a Juan de la Cierva Postdoctoral Fellowship, project JDC2022-048934-I, funded by MCIN/AEI/10.13039/501100011033 and by the European Union “NextGenerationEU/PRTR”, and a Humboldt Research Fellowship funded by the Alexander von Humboldt
foundation. V.M.R. aknowledges support through grant RYC2020-029387-I funded by MICIU/AEI/10.13039/501100011033 and by "ESF, Investing in your future", and from the Consejo Superior de Investigaciones Cient{\'i}ficas (CSIC) and the Centro de Astrobiolog{\'i}a (CAB) through the project 20225AT015 (Proyectos intramurales especiales del CSIC). V.M.R. and D.S.A. acknowledge support from the grant CNS2023-144464 funded by MICIU/AEI/10.13039/501100011033 and by “European Union NextGenerationEU/PRTR”. 
D.S.A. acknowledges the financial support provided by the Comunidad de Madrid through the Grant PIPF-2022/TEC-25475. L.C. acknowledges support from a research fellowship from the ”la Caixa” Foundation (ID 100010434 -- fellowship code LCF/BQ/PR25/12110012). B.T. acknowledges Spanish Ministry of Science support from grants PID2022-137980NB-100 and PID2023-147545NB-I00. A.L-G. and D.S.A acknowledge support from the Consejo Superior de Investigaciones Cient{\'i}ficas (CSIC) and the Centro de Astrobiolog{\'i}a (CAB) through the project 20225AT015 (Proyectos intramurales especiales del CSIC). E.J.C. acknowledges support from the Basque Government (project IT1491-22) and from grant PID2023-147698NB-I00 funded by MCIN/AEI/10.13039/501100011033 and ERDF/EU, and the CSIC I-LINK project ILINK25125.


\begin{sidewaystable}
\caption{Derived physical parameters and modelled abundances of erythrulose and chemically-related species toward the G+0.693 molecular cloud.}\label{tab:abun}
\begin{tabular*}{\textheight}{@{\extracolsep\fill}lcccccc}
\toprule%
Molecule & $N$  & $T_{\rm ex}$ & $v_{\rm LSR}$ & FWHM & $\chi_{\rm obs}$$^a$ & $\chi_{\rm mod}$$^d$\\
& ($\times$10$^{13}$ cm$^{-2}$) & (K) & (km s$^{-1}$) & (km s$^{-1}$) & ($\times$10$^{-10}$) & ($\times$10$^{-10}$) \\
\midrule
Glycolaldehyde (HOCH$_2$CHO)$^b$ &	9.3$\pm$0.3 &	21.8$\pm$0.8 &	68.9$\pm$0.2 &	19.4$\pm$0.5 &	6.9$\pm$0.2 & $\sim$20 \\
aGg'-Ethylene Glycol [aGg'-(CH$_2$OH)$_2$] &	23.5$\pm$5.2	& 10.3$\pm$2.6 &	67.9$\pm$2.8 &	21$^c$	& 17.4$\pm$3.9 & $\sim$100 \\
Glyceraldehyde (HOCH$_2$CH(OH)CHO) & $\leq$0.5	& 10$^c$	& 69$^c$ &	21$^c$ & $\leq$0.4 & $\sim$20 \\
Dihydroxyacetone (HOCH$_2$COCH$_2$OH) & $\leq$1.0	& 10$^c$	& 69$^c$ & 21$^c$ & $\leq$0.7 & $\sim$50 \\
Glycerol (HOCH$_2$CH(OH)CH$_2$OH) & $\leq$2.5	& 10$^c$	& 69$^c$ &	21$^c$ & $\leq$1.9 & $\dots$ \\
Erythrulose (C$_4$H$_8$O$_4$) &	8.7$\pm$0.8	& 11.3$\pm$1.8	& 69$^c$ & 22$^c$ &	6.4$\pm$0.6 & $\sim$20 \\
\botrule
\end{tabular*}
\footnotetext[a]{Molecular abundances calculated using $N$(H$_2$) = 1.35$\times$10$^{23}$cm$^{-2}$ (ref.\cite{martin2008}).}
\footnotetext[b]{Derived physical parameters taken from ref.\cite{Rivilla2022a}.}
\footnotetext[c]{This parameter was fixed in \textsc{Madcuba} SLIM-\textsc{Autofit} to reach convergence.}
\footnotetext[d]{Modelled abundances at 1.6 Myr from the simulation with $\zeta$=1.3$\times$10$^{14}$ s$^{-1}$.}
\end{sidewaystable}    

\begin{table}[ht]
\caption{{\bf Observed and predicted abundance ratios with respect to erythrulose.} The predicted abundances are taken from the model with $\zeta \geq$ 1.3$\times$10$^{-14}$ s$^{-1}$ at $\sim$1.6 Myr (Table \ref{tab:abun}).}\label{tab:abun2}%
\begin{tabular}{@{}lccc@{}}
\toprule
Abundance Ratio & Observed & Modelled & $\frac{\rm Observed}{\rm Modelled}$ \\
\midrule
Glycolaldehyde / Erythrulose & 1.1 & $\sim$1 & $\sim$1 \\
Ethylene Glycol / Erythrulose & 2.7 & $\sim$5 & $\sim$0.5 \\
Erythrulose / Glyceraldehyde & 17.4 & $\sim$1 & $\sim$17 \\
Erythrulose / Dihydroxyacetone & 8.7 & $\sim$0.4 & $\sim$22 \\
\botrule
\end{tabular}
\end{table}







\newpage

\section*{Extended Data}

\begin{sidewaystable}
\caption{{\bf Spectroscopic information of the erythrulose transitions shown in Figure 1.} The S/N ratio is computed from the integrated signal ($\int$ $T$$\mathrm{_A^*}$d$v$) and its associated noise level, $\sigma$ = rms $\times$ $\sqrt{\delta v \times \mathrm{FWHM}}$, where $\delta$$v$ is the velocity resolution of the spectra and the FWHM is fitted from the data. We also provide the contribution of the contamination to the overall area after subtracting the LTE fit of erythrulose from the observed spectrum over a velocity range of $\pm$FWHM (Residual area in \%). We consider that the observed feature is mostly unblended if the residual area is $\leq$25\%. A similar approach has been adopted in recent detections of complex organic molecules in line rich sources\cite{Araki2026}. Note $^{a}$ indicates that the line is considered blended although the residual area is $\leq$25\%, because of a close (CH$_2$OH)$_2$ line. Values in parenthesis in the first column correspond to frequency uncertainties given in units of MHz.}
\label{tab:spec}
\begin{tabular*}{\textheight}{@{\extracolsep\fill}lcccccccc}
\toprule
Frequency &	Transition	& Log I	& E$_{low}$	& rms & $\int$ $T$$\mathrm{_A^*}$d$v$ & S/N & Residual & Blending \\
(MHz) &	($J',K'_{-1},K'_{+1}\rightarrow J'',K''_{-1},K''_{+1}$) &	(nm$^{2}$MHz) &	(cm$^{-1}$) & (mK) & (mK km s$^{-1}$) & & & \\
\midrule
{\bf 42023.08 (0.12)} & 7,7,0$\rightarrow$6,6,0 & -6.339& 4.0 & 0.4 & 81.9 & 32 & 25\% & {\bf mostly unblended} \\
{\bf 42023.08 (0.12)}	& 7,7,1$\rightarrow$6,6,1 &	-6.339 & 4.0 & 0.4 & & & & \\ \hline
44611.73 (0.12) & 8,7,1$\rightarrow$7,6,1 & -6.279 & 4.6 & 0.6 & 81.6 & 22 & 37\% & blended (c-C$_6$H$_5$CN) \\
44611.73 (0.12) & 8,7,2$\rightarrow$7,6,2	& -6.279 & 4.6 & 0.6 & & & & \\ \hline	
{\bf 40931.91 (0.07)} & 8,6,2$\rightarrow$7,5,2 & -6.407 &	4.0 & 0.5 & 72.7 & 23 & 13\% & {\bf mostly unblended} \\
{\bf 40932.25 (0.07)} & 8,6,3$\rightarrow$7,5,3 & -6.407 & 4.0	& 0.5 & & & \\ \hline
38345.96 (0.07) & 7,6,1$\rightarrow$6,5,1 & -6.478 & 3.3 & 0.3 & 71.4 & 36 & 40\% & blended HCCCH$_2$CN \\
38346.02 (0.07) & 7,6,2$\rightarrow$6,5,2	& -6.478 & 3.3 & 0.3 & & & & \\ \hline	
{\bf 35442.28 (0.06)} & 15,1,15$\rightarrow$14,1,14 & -6.274 &	8.4 & 0.4 & 66.8 & 24 & 20\% & {\bf mostly unblended} \\  
{\bf 35443.77 (0.06)} & 15,0,15$\rightarrow$14,0,14	& -6.274 & 8.4 & 0.4 & & & \\	\hline
{\bf 39782.32 (0.04)}	& 9,5,4$\rightarrow$8,4,4 &	-6.462	& 4.1 &	0.3 & 32.4 & 17 & 8\% & {\bf mostly unblended} \\ \hline
34639.84 (0.04) & 14,1,3$\rightarrow$13,1,12 &	-6.333 &	7.9 &	0.5 &	31.9	& 9 & 35\% &	blended (U-line) \\ \hline
32227.63 (0.03)$^a$ &	13,2,12$\rightarrow$12,2,11	& -6.427 &	6.8	& 0.5 & 31.7 & 9 & 20\% & blended (CH$_2$OH)$_2$ \\ \hline
{\bf 40329.80 (0.07)} & 15,3,12$\rightarrow$14,3,11 &	-6.180 &	9.9	& 0.5 & 30.2 & 9 & 24\% & {\bf mostly unblended} \\ \hline
{\bf 33385.03 (0.03)} & 13,3,11$\rightarrow$12,3,10 &	-6.408 &	7.3	& 0.5 & 30.1 & 9 & 18\% &  {\bf mostly unblended} \\ \hline
39073.86 (0.05) &	15,4,12$\rightarrow$14,4,11 &	-6.222 &	10.1 & 0.4 &	27.6 & 10 & 49\% & blended (U-line) \\ \hline
37059.88 (0.02) &	10,3,8$\rightarrow$9,2,8 &	-6.635 &	4.1 & 0.4 &	23.6 & 9 & 52\% & blended (U-line) \\ \hline
\botrule
\end{tabular*}
\end{sidewaystable}

\section*{Supplementary Information}

\begin{figure}[htb!]
\centering
\includegraphics[width=\hsize]{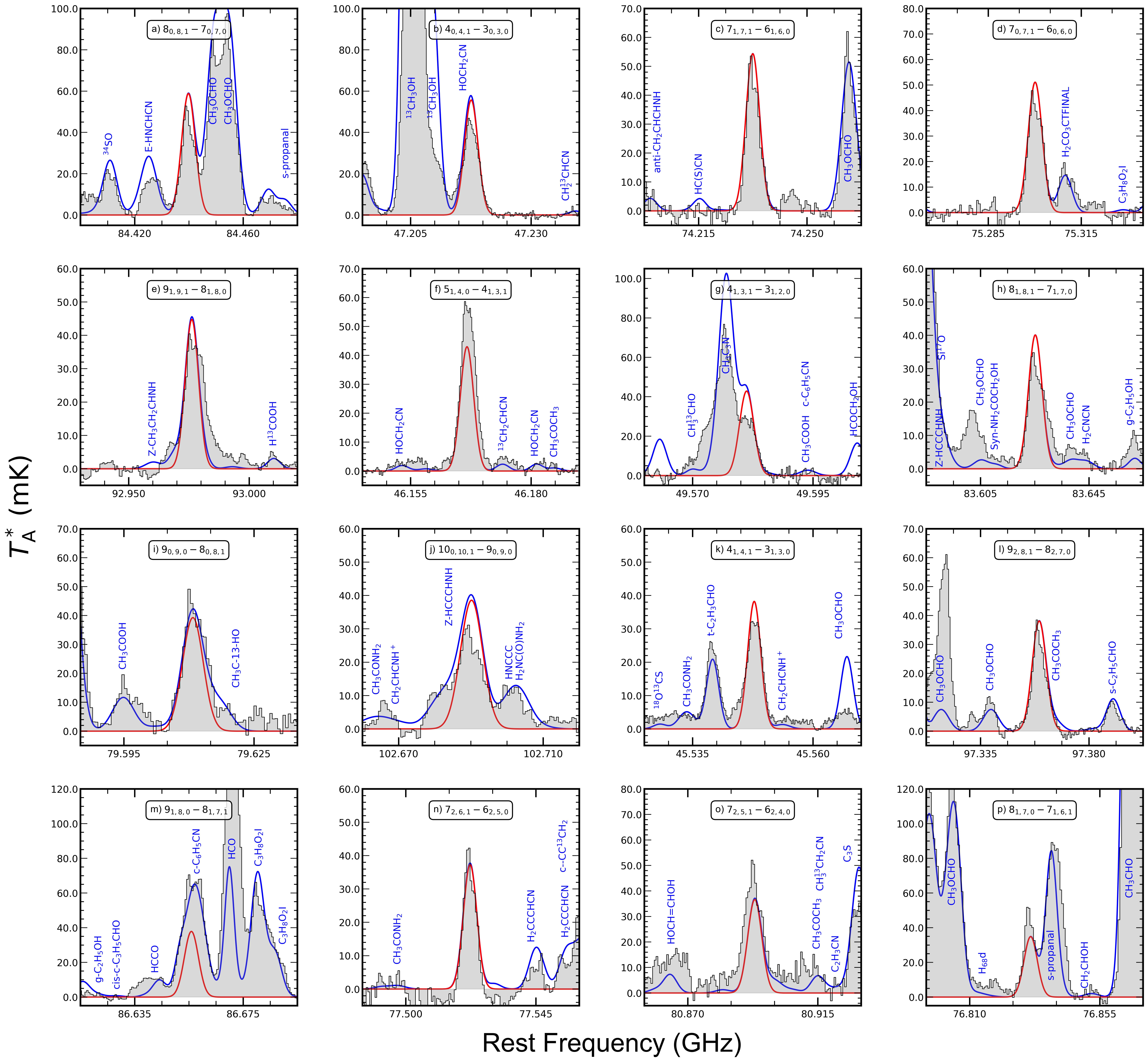}
\caption{{\bf Representative clean and slightly blended transitions of the lowest energy conformer of ethylene glycol, the aGg' conformer [aGg'-(CH$_2$OH)$_2$].} Red lines show the contribution of the individual aGg'-(CH$_2$OH)$_2$ lines, while the blue lines represent the predicted spectra after considering the emission from all molecular species identified toward G+0.693. The quantum numbers of the transitions are shown in the upper part of each panel. In blue, we also report other molecular species contributing to the total observed spectra in the vicinity of the aGg'-(CH$_2$OH)$_2$ lines.}
\label{fig:glycol}
\end{figure}

\bmhead{LTE Analysis of aGg'-Ethylene Glycol [aGg'-(CH$_2$OH)$_2$] with MADCUBA}
As discussed in the main text, one of the possible precursors of erythrulose is ethylene glycol [or (CH$_2$OH)$_2$]. Ethylene glycol has several conformers, for which the aGg' conformer is the lowest in energy. Figure \ref{fig:glycol} and  Table \ref{tab:spec-glycol} report the spectra and the spectroscopic information of representative transitions of aGg'-(CH$_2$OH)$_2$. As seen from Figure \ref{fig:glycol}, aGg'-(CH$_2$OH)$_2$ is clearly detected with many clean (or slightly blended) bright transitions (peak intensities $\geq$30 mK). The physical parameters of aGg'-(CH$_2$OH)$_2$ derived using SLIM-AUTOFIT are reported in Table 1 of the main text.

\begin{sidewaystable}
\caption{{\bf Spectroscopic information of the aGg'-(CH$_2$OH)$_2$ transitions shown in Figure \ref{fig:glycol}}.}\label{tab:spec-glycol}
\begin{tabular*}{\textheight}{@{\extracolsep\fill}lcccc}
\toprule
Frequency &	Transition	& Log I	& E$_{low}$	& Blending \\
(GHz) &	($J',K_{a}',K_{c}',v'\rightarrow J'',K_{a}'',K_{c}'',v''$) &	(nm$^{2}$MHz) &	(cm$^{-1}$) & \\
\midrule
84.439522 &	8,0,8,1$\rightarrow$7,0,7,0	& -4.690 &	9.3 &	clean \\
47.217410 &	4,0,4,1$\rightarrow$3,0,3,0	& -5.439 &	2.0	& clean \\
74.232243 &	7,1,7,1$\rightarrow$6,1,6,0	& -4.921 &	7.1	& clean \\
75.299876 &	7,0,7,1$\rightarrow$6,0,6,0 & -4.947 &	7.0	& clean \\
92.975886 & 9,1,9,1$\rightarrow$8,1,8,0	& -4.600 &	11.9 &	clean \\
46.166540 &	5,1,4,0$\rightarrow$4,1,3,1 & -5.440 &	4.1	& slightly blended (U-line) \\
49.581050 & 4,1,3,1$\rightarrow$3,1,2,0 & -5.502 &	2.5 & slightly blended (CH$_3$C$_3$N) \\
83.624924 &	8,1,8,1$\rightarrow$7,1,7,0	& -4.857 &	9.4	& clean \\
79.610634 &	9,0,9,0$\rightarrow$8,0,8,1	& -4.733 &	12.1 & slightly blended (CH$_3$$^{13}$CHO) \\
102.689838	& 10,0,10,1$\rightarrow$9,0,9,0	& -4.442 &	14.7 &	slightly blended (Z-HCCCHNH \& HNC$_3$) \\
45.547640 &	4,1,4,1$\rightarrow$3,1,3,0	& -5.605 &	2.3	& clean \\
97.359028 &	9,2,8,1$\rightarrow$8,2,7,0	& -4.548 &	13.6 &	clean \\
86.655611 &	9,1,8,0$\rightarrow$8,1,7,1	& -4.634 &	13.3 & blended (c-C$_6$H$_5$CN) \\
77.521974 & 7,2,6,1$\rightarrow$6,2,5,0	& -4.980 &	8.5	& clean \\
80.892911 &	7,2,5,1$\rightarrow$6,2,4,0	& -4.957 &	8.6	& clean \\
76.830836 &	8,1,7,0$\rightarrow$7,1,6,1	& -4.899 &	10.5 &	slightly blended (s-propanal) \\
\botrule
\end{tabular*}
\end{sidewaystable}

\begin{figure}[htb!]
\centering
\includegraphics[width=1.0\hsize]{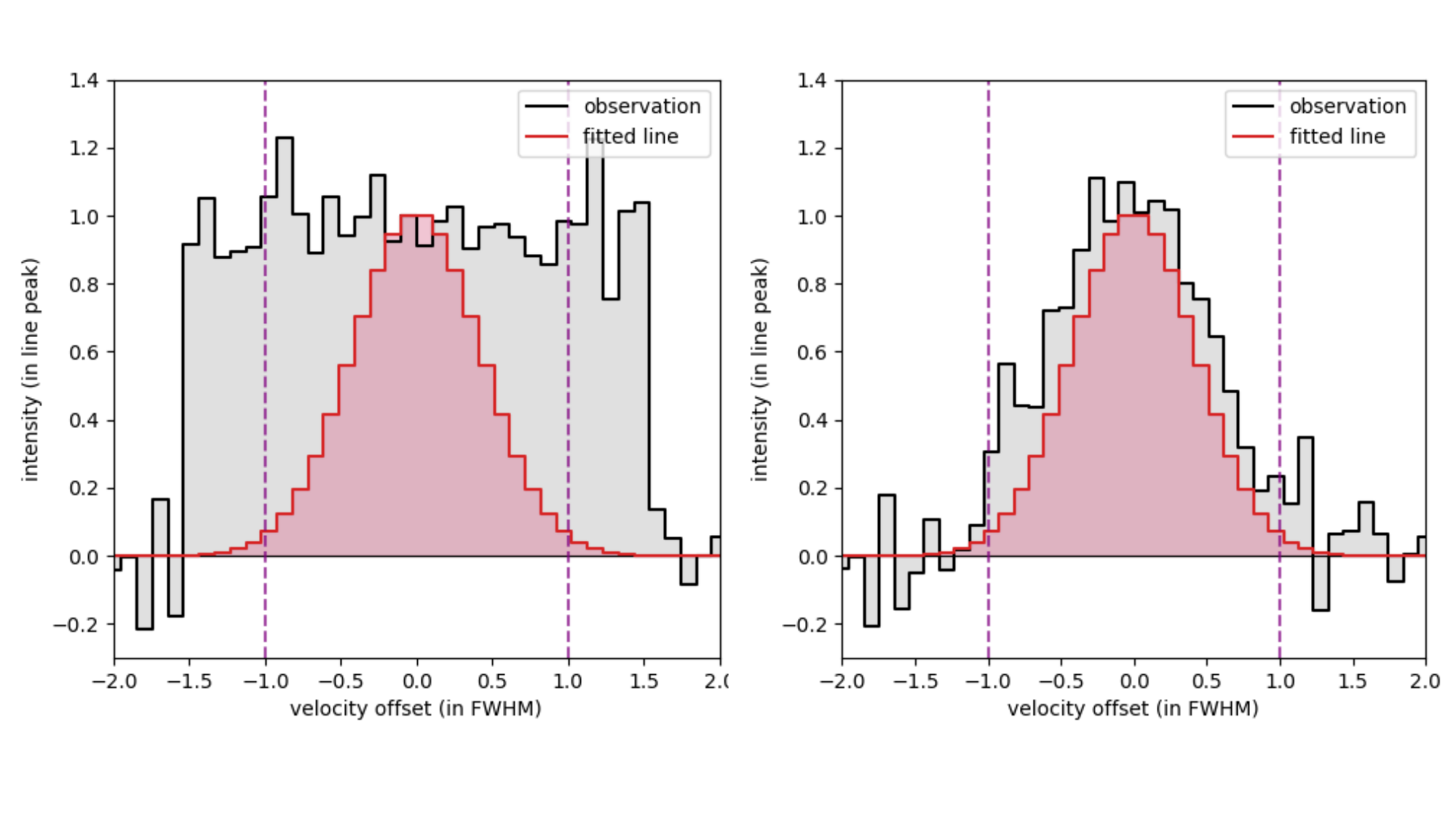}
\caption{\textbf{Comparison between mock spectra (black histograms and grey shaded areas) and Gaussian line profiles (red histograms and red shaded areas).} Left panel shows the most extreme case of blending with a mock flat spectrum, while the right panel shows the case of minor blending contribution. The noise in the mock spectra corresponds to 10\% the peak value of the Gaussian line profile. Dashed vertical lines indicate the $\pm$FWHM velocity range over which 98.2\% of the area underneath the Gaussian profile is considered.}
\label{fig:gaussian}
\end{figure}

\bmhead{Threshold to establish the level of line blending}
As described in Methods, the criteria to establish whether a line is mostly-unblended or blended is that the residuals obtained after subtracting the contribution of the erythrulose LTE line fits to the observed spectra are $\leq$25\%. To demonstrate the validity of this threshold, we have performed numerical simulations of mock spectra with flat line profiles (representing the most extreme case of heavy line blending) and with slightly-broader Gaussian line profiles with a linewidth = 1.3 $\times$ the original linewidth and a peak intensity = 1.1 $\times$ the original peak intensity (representing minor blending contribution; Figure \ref{fig:gaussian}). The residuals are calculated by subtracting the Gaussian line (red shaded) area over a velocity range of $\pm$FWHM (or 2$\times$FWHM), covering 98.2\% of the total area of the Gaussian profile. For the flat spectrum case (left panel in Figure \ref{fig:gaussian}), the residuals amount to $\sim$50\%, while for the slightly-blended case (right panel), they are $\leq$25\%, respectively. This justifies our choice of $\leq$25\% as the threshold for classifying lines as mostly-unblended. 

\bmhead{Modeling of amorphous solid water (ASW) ice}
Our mechanistic approach for the formation of erythrulose involves the non-diffusive association of activated glycolaldehyde (g) and ethylene glycol (e) on ASW ice. Non-diffusive chemistry has been invoked to explain the formation of COMs under cold ISM conditions, and it is especially important in sources where the temperature of the dust remains below 30 K\cite{Lamberts2019,Simons2020}. Therefore, the first step is to obtain the complex formed between both species (g-e) adsorbed on the ice surface. To simulate the ASW ice, we build a cluster of 100 water molecules starting by solvating an optimized water molecule with GFN2-xTB\cite{gfn2} through an Automated Molecular Cluster Growing calculation\cite{qcg} implemented in the CREST\cite{crest} program at the same level of theory mentioned above. This cluster was optimized and used as the initial structure for performing molecular dynamics. The cluster of water molecules was encapsulated in a sphere of 27 Bohr radius to prevent evaporation of the water molecules, and a molecular dynamics simulation was carried out for 100 ps at 300 K using the GFN-FF force field.\cite{gfnff} The structure of the cluster was extracted every 10 ps and used as input for propagating another molecular dynamics simulation for 10 ps at a temperature of 10 K. These 10 clusters were optimized using the GFN2-xTB method, and their electronic energies were refined at the DFT level using the PW6B95(D3)\cite{Zhao2005} method in combination with the def2-TZVP basis set.\cite{Ahlrichs2005} Among all these clusters, the one with the lowest relative energy compared to the least stable cluster was selected, in this case -100.0 kJ mol$^{-1}$.

\begin{figure}[htb!]
\centering
\includegraphics[width=\hsize]{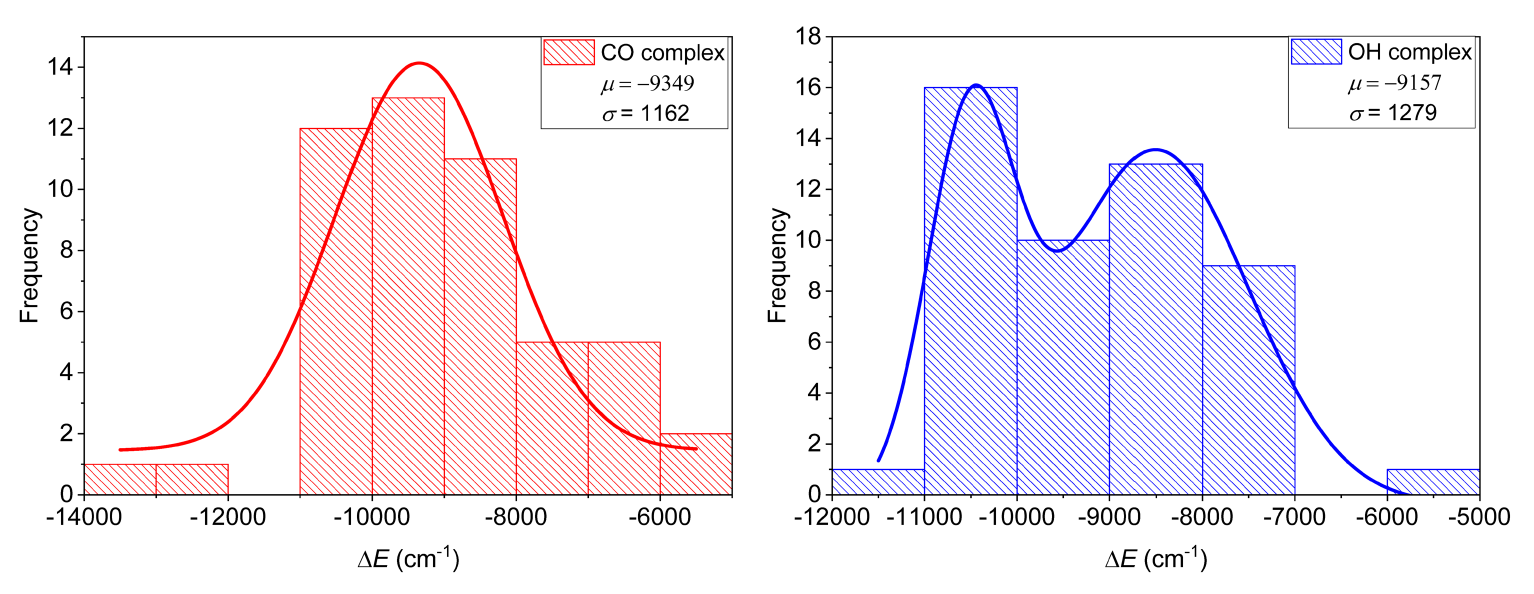}
\caption{\textbf{Binding energies of the g-e CO (red) and OH (blue) complexes on the ASW.} Solid lines represent gaussian fits to the binding energies.}
\label{fig:Ebind}
\end{figure}

\bmhead{Generation of the glycolaldehyde-ethylene glycol complex (g-e) in ASW}
In our simulations, we assume that glycolaldehyde and ethylene glycol are formed close to each other on the ice surface. As glycolaldehyde and ethylene glycol are molecules with very polar groups, the aldehyde oxygen and the hydroxyl groups will point towards the water ice. With these premises, we placed several glycolaldehyde molecules around ethylene glycol with all polar groups pointing in the same direction. Each pair was optimized, resulting in only two possible structures: one structure in which the carbonyl group of glycolaldehyde interacts with a hydroxyl group of ethylene glycol (referred to as CO), and a second structure in which the two hydroxyl groups of the pair interact (OH). These two structures were optimized and used as fragments in a rigid docking study (aISS)\cite{aiss}, obtaining glycolaldehyde-ethylene glycol complexes (g-e) placed in different areas of the ice. For each complex, a sampling of 50 interactions was performed. All energies of the g-e complex on the ASW were refined at the aforementioned DFT level. The binding energies of the CO and OH complexes to the ASW (Figure \ref{fig:Ebind}) were very similar (9349 $\pm$ 1162 cm$^{-1}$ and 9157 $\pm$ 1279 cm$^{-1}$, respectively). However, the highest interaction for CO was 13264 cm$^{-1}$, and for OH, it was 11014 cm$^{-1}$. Then we selected the structure with the highest binding energy of the CO complex for the mechanistic study. Note that to calculate the binding energies, we used the energy of all fragments separately as a reference and are not ZPE-corrected because these attractive interactions are high in energy and the ZPE does not significantly alter the results.

\bmhead{Electronic structure calculations}
All calculations carried out to study the reaction mechanism were done with the Gaussian 16 package\cite{gaussian}. For the association of the two fragments to occur in the ASW, first, the molecules must be activated, i.e. they must be radicals. The activation can occur after H abstraction reactions in both fragments. Hydrogen abstraction reactions are known to take place in ASW ice under interstellar conditions as demonstrated by laboratory experiments and astrochemical simulations\cite{fedoseev2017,Ioppolo2021,garrod2022,Borshcheva2025}. The only hydrogens that can be abstracted in both fragments are those pointing out of the ASW ice, which correspond to C-H bonds (see optimized geometries in Figure 2). For this calculation, we used a three-layer ONIOM where the CO complex and the first solvation layer were treated at DFT level. The last water molecules included in the DFT level calculations interact with another layer of water molecules treated at a medium level of theory using the semiempirical PM7R6 method\cite{pm7}. The rest of the water molecules, already away from the reaction area, were treated with the UFF force field\cite{uff}, resulting in a PW6B95(D3)/def2-TZVP:PM7R6:UFF level of theory. All optimizations were carried out without restrictions, and the stationary points were characterized by frequency calculations at the same level of theory, showing no imaginary frequencies for energy minima and saddle points, respectively. For the calculation of the ZPE, a scale factor of 0.974 was used\cite{freqscal}. The harmonic approximation was used for frequency calculations, which introduces notable errors in entropy for normal modes of very low magnitudes. Therefore, all frequencies below 50 cm$^{-1}$ were fixed at this value. 

After the first H abstraction reaction, the complex adsorbed on the ice is a doublet (g*-e), so subsequent incident hydrogen atoms can approach the complex with the spin up or spin down, giving rise to two possible spin states, a singlet diradical and a triplet, respectively. The wavefunction for the high spin state was calculated and used as a reference for performing the calculation of the low spin state wavefunction (singlet diradical). In no case did we find transition states for the second H abstraction for the singlet diradical state. All transition states found for the second H abstraction are for triplet states. 

Once the g*-e* complex is formed, which is in the triplet state, it must change the spin of one of the carbons to form erythrulose. Therefore, the singlet-triplet ISC transition was considered. The spin-orbit coupling was calculated at the complete active self-consistent field\cite{casscf} with an active space of six electrons in five molecular orbitals, leaving a CASSCF(6e,5o)/def2-TZVP level of theory. These calculations were carried out with the ORCA software\cite{orca}. To do this, we removed all the water molecules treated with the UFF force field and the ones that were treated with the PM7R6 method were calculated with the GFN2-xTB.

\bmhead{Kinetic calculations}
Since all polar groups in the g-e complex are oriented towards the ice surface, we focused on the abstraction of hydrogen atoms that remain non-interacting with the surface. To obtain the relative energies of the transition states, an Intrinsic Reaction Coordinate (IRC) calculation was performed to obtain the geometry of the reactant complex, which was taken as a reference to obtain the energy barriers and reaction energies. The rate constants were obtained using transition state theory (TST) using the Pilgrim software\cite{pilgrim}:

\begin{equation}
k(T) = \kappa \frac{k_B T}{h} \frac{Q_\text{rot}^\text{SP} Q_\text{vib}^\text{SP} Q_\text{elec}^\text{SP}}{Q_\text{rot}^\text{C} Q_\text{vib}^\text{C} Q_\text{elec}^\text{C}} e^{\frac{-V^\text{SP}}{k_B T}}
\end{equation}

where $k_{B}$ and $h$ are the Boltzmann and Plank constants, respectively; $Q$ refers to the partition function with the super index C and SP corresponding to the complex and saddle point, respectively; and $V^{SP}$ is the potential energy of the saddle point. Since we are computing a mechanism on a surface, the rotational partition functions $Q_{rot}^SP$  and $Q_{rot}^C$ were set to 1. The rotational partition function was not fixed for molecular hydrogen since the relative energy of the products with respect to the complex was obtained with respect to the energy asymptote of the exit channel. Else, the rotational partition function was set to unity for the newly formed radical, as was done for the complex and the saddle point. The classical rate constants obtained with TST were corrected by multiplying them by the tunneling transmission coefficient ($\kappa$) using the Eckart approximation\cite{eckart1930penetration}. The final step of the reaction, in which the singlet diradical recombines to form erythrulose, follows a Langmuir–Hinshelwood mechanism for which we could not calculate rate constants since we could not characterize any transition structure connecting the two minima. To approximate the ISC rate in the diradical intermediate g*-e* we use Marcus' semi-classical theory, given by\cite{Marcus1964,Marcus1985}

\begin{equation}
k_\text{ISC}(T) = \frac{2\pi}{h} V_\text{SOC}^2 \sqrt{\frac{\pi}{\lambda k_B T}} e^{\frac{-(\lambda + \Delta G)^2}{4\lambda k_B T}}
\end{equation}

where $\Delta G$ is the free energy difference between the two spin states for their equilibrium geometries, $\lambda$ is the reorganization energy, which was calculated following the four-point technique proposed by Nelsen\cite{Nelsen1987} and $V_{SOC}$ is the spin-orbit coupling between the two spin states.

\bmhead{Further details on the reaction mechanism between glycolaldehyde and ethylene glycol on ASW} 
In the main text we conclude that reaction 3-g-e is the most favourable reaction on ASW because its energy barrier (4.1 kJ mol$^{-1}$) is significantly lower than those of the H abstraction reactions 1-g-e, 4-g-e, 5-g-e or 2-g-e (see panel {\bf a} in Figure 2). Whereas reactions 1-g-e, 4-g-e and 5-g-e present, respectively, energy barriers of 11.5, 22.2 and 14.9 kJ mol$^{-1}$, the H addition to the C atom in reaction 2-g-e is 23.4 kJ mol$^{-1}$. Therefore, all these reactions cannot compete with reaction 3-g-e based on their classical energy barriers. Although reaction 1-g-e has a classical energy barrier just 7.4 kJ mol$^{-1}$ above that of the reaction 3-g-e, the rate constant is significantly lower because the imaginary frequency of the saddle point is 289.4 cm$^{-1}$ lower than that found for reaction 3-g-e, which leads to a very significant decrease in the transmission coefficient (see below).

When the activated complex g*-e is formed after the initial H abstraction reaction 3-g-e (see Figure 2, panel {\bf a}, in the main text), it adopts a doublet spin state. Subsequent collisions with additional H atoms can then result in the formation of two possible spin states. When the system is in a triplet state, the approach of H towards the activated carbon is repulsive, while collisions with the available H atoms from g*-e lead to further H abstractions, forming H$_2$ and a doubly activated complex in a triplet state (g*-e*). We note that H atoms with antiparallel spin approaching the activated carbon of the aldehyde could revert to the inactivated g-e complex, as the potential between these two atoms is attractive and barrierless. This reversibility to the g-e complex could hinder the reaction process or lead to the desorption of the reactants from the ice, given that hydrogenation of the activated aldehyde carbon is exothermic by 356.8 kJ mol$^{-1}$. Nevertheless, it is well known that collisions with atomic hydrogen on molecules within ice tend towards more complex systems through diradical recombination. This is due to the different orientation of the new incident H atoms in subsequent collisions added to the orientation of the COMs on the ASW, which allows the reaction to proceed through more H abstraction reactions\cite{Lamberts2019,Simons2020}. In our case, the incoming H atom towards hydrogen (3) of ethylene glycol in the g*-e complex (Figure 2, panel {\bf b}) has a different orientation from that required to collide with the activated carbon of g*-e, yielding the g*-e* complex shown in Figure 2 (panel {\bf c}) in an extremely fast process. 

\begin{figure}[htb!]
\centering
\includegraphics[width=\hsize]{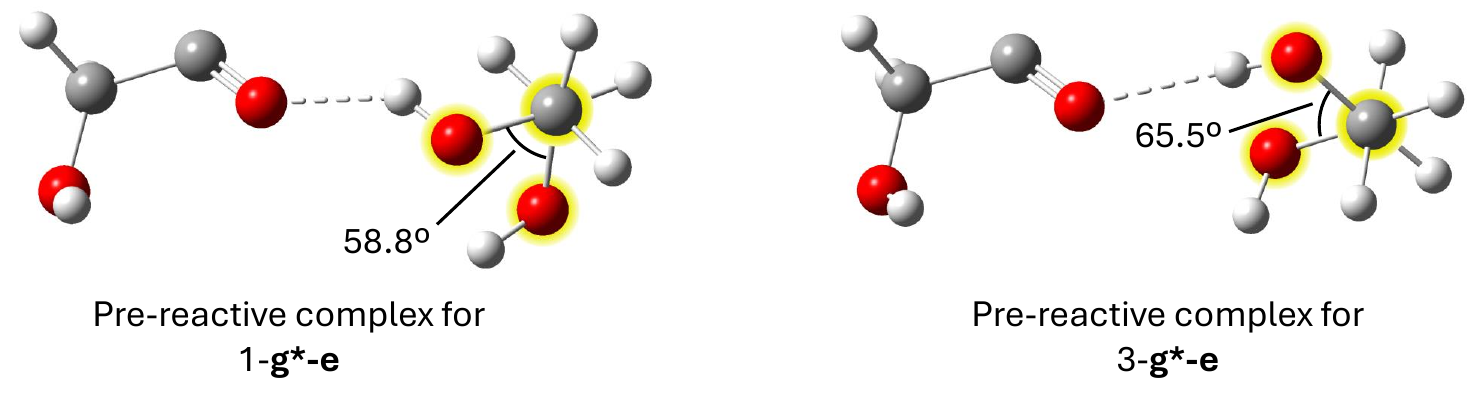}
\caption{\textbf{Optimized geometries of the pre-reactive complexes for reactions 1-g*-e and 3-g*-e.} For the shake of clarity, water molecules and the H atom have been removed.}
\label{fig:geometries}
\end{figure}

As shown in Figure 2 (panel {\bf b}), there are three highlighted hydrogen atoms in the g*-e complex available to react with colliding H when the system is in the triplet state. However, our calculations show that the formation of the pre-reactive complex for the abstraction reaction 1-g*-e is repulsive, being 3.0 kJ mol$^{-1}$ above the reactants' asymptote. In contrast, the pre-reactive complex for the reaction 3-g*-e is 25.0 kJ mol$^{-1}$ below the reactants' asymptote. Therefore, the energy difference between these two complexes is 28.0 kJ mol$^{-1}$, which makes reaction 1-g*-e unfeasible under ISM conditions. The inspection of both complexes shows that the interaction energy of g*-e with H at two different positions is not the reason for this large energy difference. The geometries of the pre-reactive complexes for hydrogen (1) and hydrogen (3) without considering the incoming hydrogen are also very similar, showing a geometry RMSD deviation of 1.5 \AA. The main difference between the two geometries lies in the orientation of ethylene glycol on the ASW ice, which slightly rotates its position along the C-C bond (see Figure \ref{fig:geometries}). This small reorientation induces a change in the O-C-C-O dihedral angle, being 58.8$^\circ$ and 65.5$^\circ$ for the least stable and the most stable complex, respectively. From an electronic perspective, the greater the distance between the oxygen atoms of these two hydroxyl groups, the lower the steric repulsion. The change in orientation decreases the directionality of the hydrogen bond with respect to the carbonyl group, which destabilizes the system.

The H abstraction reaction 2-g*-e forms hydroxyketene through a hydroxyketene-ethylene glycol complex, while reaction 3-g*-e yields the doubly activated complex g*-e*. These two reactions show almost identical rate constants, with the 3-g*-e reaction being slightly favorable as can be seen from the branching ratios (Figure 2, panel {\bf g}). The energy barriers for the reactions 2-g*-e and 3-g*-e are 12.7 and 15.6 kJ mol$^{-1}$ and the reaction energies are -111.5 and -36.2 kJ mol$^{-1}$ respectively. Although the classical energy barrier favors the reaction 2-g*-e, the quantum corrected rate constants are slightly higher for 3-g*-e since the transmission coefficients ($\kappa$) are higher for the latter. It can be noticed qualitatively by inspecting the imaginary frequency of the saddle points of reactions 2-g*-e and 3-g*-e, which are 1018.5i and 1235.3i cm$^{-1}$, respectively.

Once the g*-e* complex is formed in the triplet state, one of the electron spin must change to form erythrulose. The ISC process depends mainly on three factors: the spin-orbit coupling between the two states (VSOC), the change in geometry (reorganization energy, $\lambda$) and the free energy difference between the singlet and triplet states. The spin-orbit coupling between these two states is very low (0.002 cm$^{-1}$), because there is no change in the orbital angular momentum after the spin changes, which makes these two states poorly electronically coupled. The change in the geometry of the singlet and triplet states is practically negligible, leading to a near-zero rearrangement energy (2.1$\times$10$^{-4}$ eV), which greatly increases the transition probability. This, coupled with the fact that both states are nearly degenerate near the absolute zero, makes the spin change very fast at low temperatures (Figure 2, panel {\bf h}). Although this process is fast, it is still slower than the two previous abstraction reactions. We note that the ISC process does not occur for the ketene-ethylene glycol complex. After the abstraction reaction 2-g*-e, a ketene is formed in the triplet state, where the C-C-O angle is 130$^\circ$, whereas the singlet state shows a C-C-O angle of 177.2$^\circ$. This geometrical change induces a considerable increase of $\lambda$ (2.8 eV), which reduces notably the probability of transition. Moreover, the free energy difference between the two states is in the range 114.5-154.6 kJ mol$^{-1}$ for T=10--300 K, making the transition non-viable. In this case, the spin-orbit coupling is higher than for the g*-e* complex, being 3.22 cm$^{-1}$. 

\begin{figure}[htb!]
\centering
\includegraphics[width=0.5\hsize]{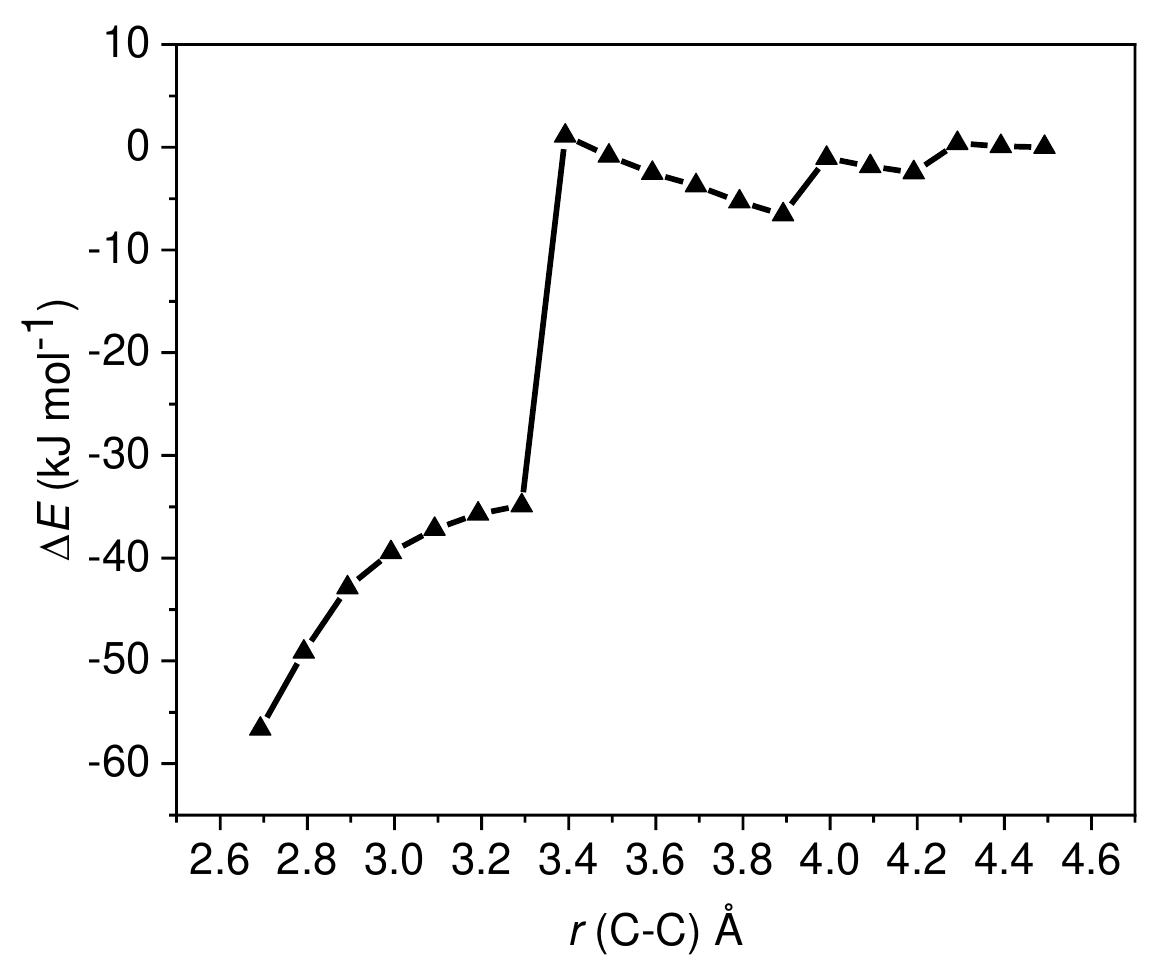}
\caption{\textbf{Relaxed scan along the C-C coordinate of the g*-e* complex.} Electronic energy change relative to the doubly activated g*-e* complex in the singlet state.  The degree of freedom scanned is the  C-C coordinate which corresponds to the two atoms holding the unpaired electrons. For each optimization along the C-C coordinate no constraints were imposed to any degree of freedom.}
\label{fig:scan}
\end{figure}

Once the doubly activated g*-e* complex is in the singlet state, recombination of the two fragments produces erythrulose. In many cases, this recombination occurs by a barrierless reaction after the second H abstraction reaction\cite{Lamberts2019,Simons2020}. However, we succeeded in characterizing the singlet diradical as an energy minimum. All attempts to characterize a saddle point leading directly to erythrulose were unfruitful, pointing to the fact that if any transition state existed between these two minima (g*-e* and erythrulose) it would be located in a very flat surface. After performing a relaxed scan by varying the C-C coordinate, we were able to detect a small energy maximum at a C-C distance of 3.4 \AA$\,$ of 1.1 kJ mol$^{-1}$ above g*-e* (Figure \ref{fig:scan}). Such an electronic energy barrier would imply a k(T) of the order of 10$^8$ s$^{-1}$. The release of the C-C degree of freedom would give rise to a more relaxed structure, decreasing the electronic energy of the system; the inclusion of ZPE would yield a lower barrier; and the tunneling effect is probably not negligible even for a wide barrier at 20 K; therefore, we can qualitatively describe this process as an extremely rapid reaction that will hardly influence the overall rate of the mechanism. Overall, the rate-determining step of the whole process is the intersystem crossing (ISC) rate of g*-e*.

\begin{figure}[htb!]
\centering
\includegraphics[width=0.9\hsize]{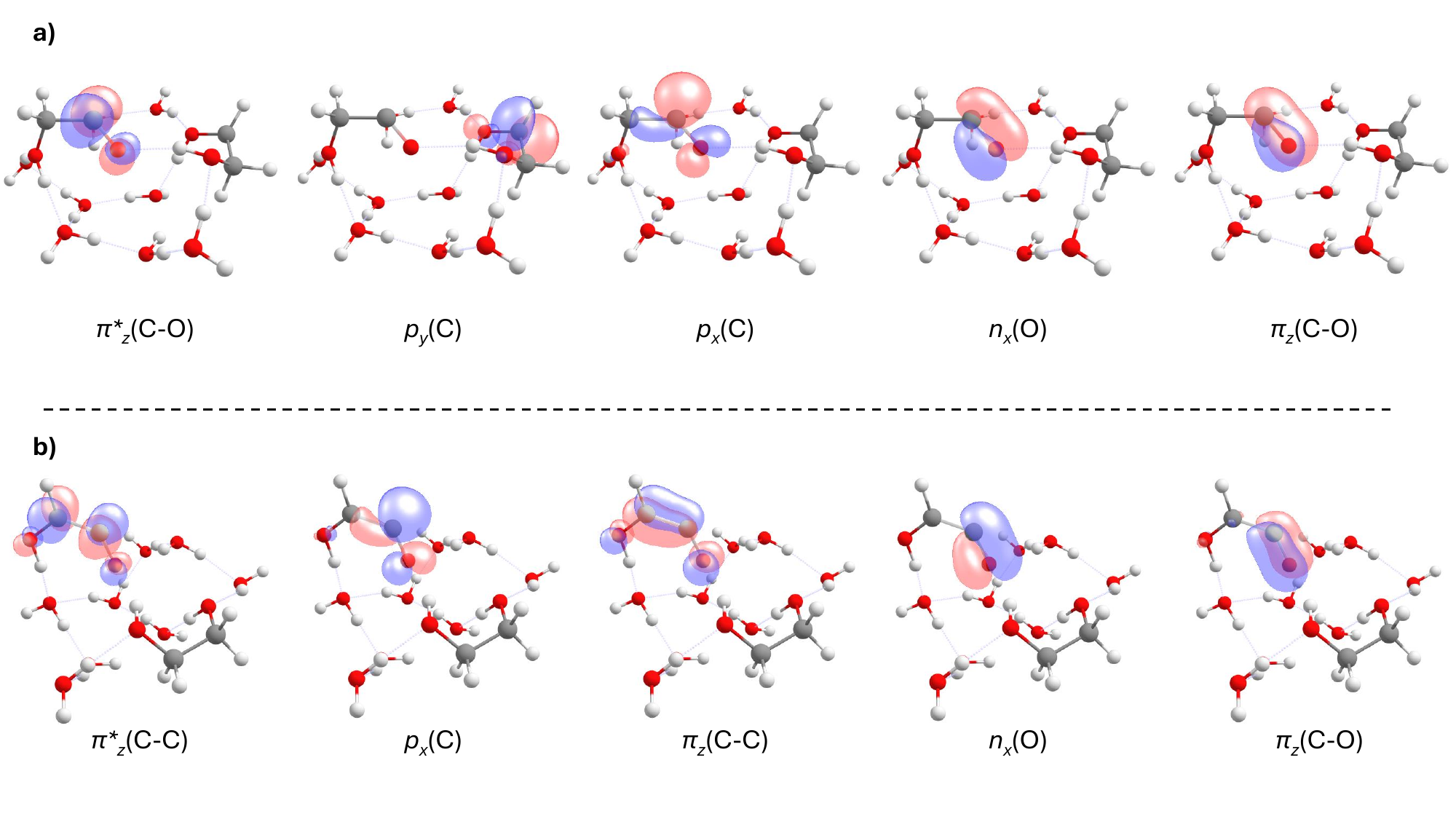}
\caption{\textbf{Active orbitals of the active space for the CASSCF(6e,5o)/def2-TZVP calculations.} Panel a) shows the active orbitals for the g*-e* complex, while panel b) for the ketene-ethylene glycol complex. Some water molecules have been removed in order to enhance the clarity of the images.}
\label{fig:orbitals}
\end{figure}

\bmhead{CASSCF active space for the g*-e* and ketene-ethylene glycol complexes}
The multiconfigurational calculations of the g*-e* and ketene-ethylene glycol complexes were carried out to derive the spin-orbit coupling. The molecular orbitals included in the active space in both complexes are the $\pi$ and $\pi^{*}$ orbital pairs, the lowest energy lone pair of the oxygen atom ($n$) as well as the $p$ type orbitals bearing the unpaired electrons (Figure \ref{fig:orbitals}). In the g*-e* complex, four out of five orbitals belong to the glycolaldehyde radical, that is, the $\pi^{*}_{z}$ and $\pi_{z}$ of the C-O bond, the lone pair of the oxygen atom ($n_{x}$) and the $p_{x}$ orbital bearing the unpaired electron. The fifth orbital of the g*-e* complex is the $p_{y}$ orbital bearing the unpaired electron of the ethyleneglycol moiety. 

The five molecular orbitals of the selected active space in the ketene-ethylene glycol complexes are in the ketene moiety. The orbitals are: The $\pi_{z}$ of the C-C bond, which bear one of the unpaired electrons, its $\pi^{*}_{z}$ pair, the $\pi_{z}$ of the C-O bond; the doubly occupied $n_{x}$ of the oxygen atom and the $p_{x}$ orbital bearing the other unpaired electron. 

\begin{figure}[htb!]
\centering
\includegraphics[width=0.85\hsize]{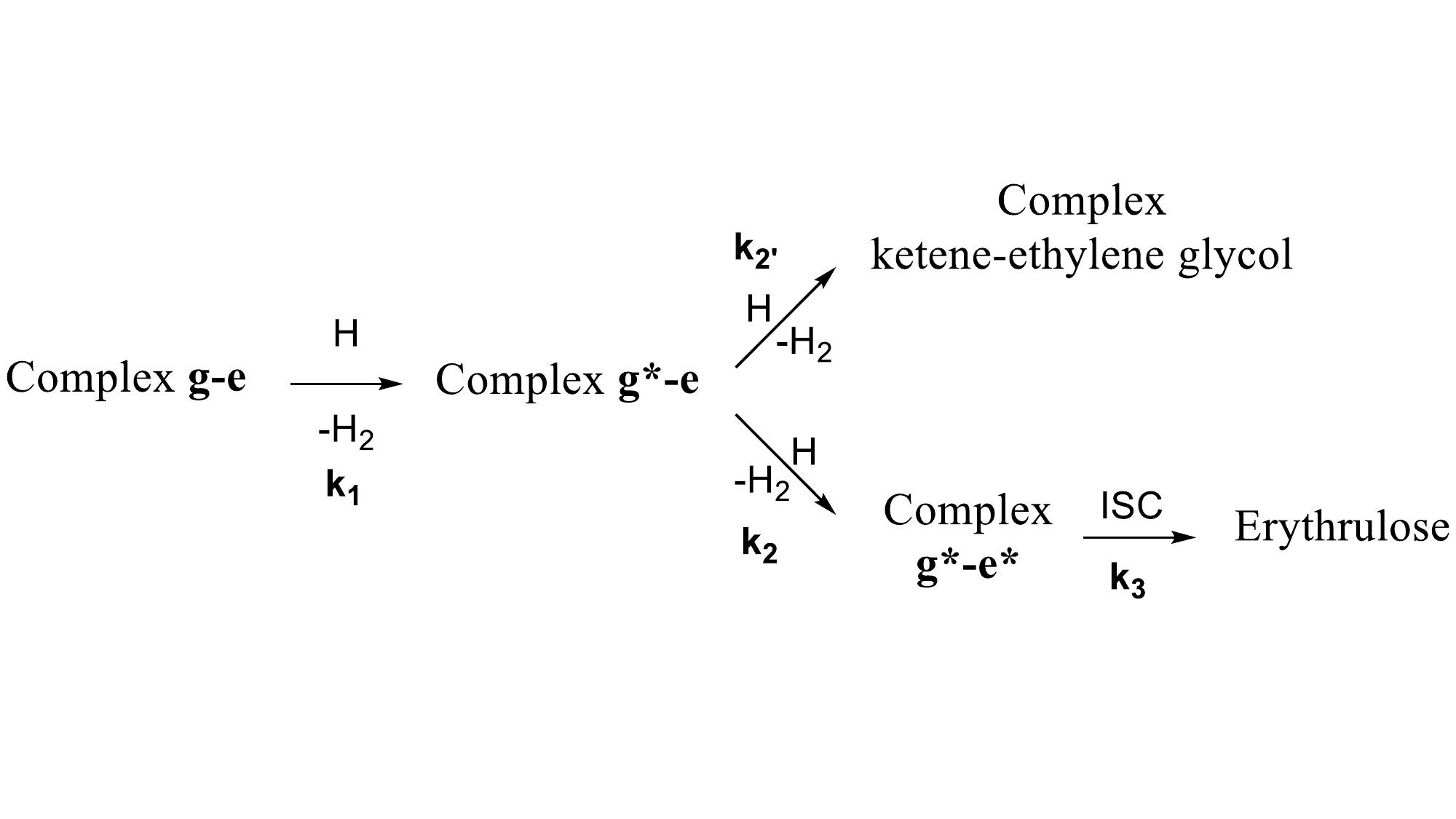}
\caption{\textbf{Summary of the mechanism of formation of erythrulose on ASW from glycolaldehyde (g) and ethylene glycol (e).} Reaction 3-g-e in Figure 2 is governed by rate constant k$_1$. Reactions 2-g*-e and 3-g*-e in Figure 2 present, respectively, rate constants k$_2$’ and k$_2$. And the ISC reaction of complex g*-e* in Figure 2 finally yields erythrulose at a rate constant k$_3$. The values of these rate constants are given in Table \ref{tab:rates} for a temperature range between 20-300 K.}
\label{fig:summary}
\end{figure}

\bmhead{Summary of the reaction mechanism}
A summary of the formation mechanism of erythrulose on ASW ice from glycolaldehyde and ethylene glycol is given in Figure \ref{fig:summary}. In Table \ref{tab:rates} we provide the rate constants of the individual H abstraction reactions and of the radical-radical recombination reaction (the ISC process) reported in Figure 2 and described in the main text. 

\begin{table}[ht]
\caption{{\bf Quantum tunneling-corrected rate constants of reactions 3-g-e (k$_1$), 2-g*-e (k$_2$') and 3-g*-e (k$_2$), and of the ISC reaction of the g*-e* complex (k$_3$).}}
\label{tab:rates}
\begin{tabular*}{\textwidth}{@{\extracolsep\fill}lcccc}
\toprule
T (K) &	k$_1$ (s$^{-1}$) &	k$_2$' (s$^{-1}$) &	k$_2$ (s$^{-1}$) &	k$_3$ (s$^{-1}$) \\
\midrule
20	& 3.08E+10	& 9.00E+06	& 9.72E+06	& 2.44E+05 \\
30	& 6.55E+10	& 1.80E+07	& 2.01E+07	& 1.04E+05 \\
40	& 1.13E+11	& 2.79E+07	& 3.22E+07	& 3.99E+04 \\
50	& 1.71E+11	& 3.85E+07	& 4.53E+07	& 1.47E+04 \\
60	& 2.35E+11	& 5.04E+07	& 5.93E+07	& 5.31E+03 \\
70	& 3.02E+11	& 6.42E+07	& 7.48E+07	& 1.91E+03 \\
80	& 3.69E+11	& 8.07E+07	& 9.18E+07	& 6.82E+02 \\
90	& 4.31E+11	& 1.01E+08	& 1.11E+08	& 2.24E+02 \\
100	& 4.91E+11	& 1.26E+08	& 1.33E+08	& 8.01E+01 \\
120	& 5.98E+11	& 1.94E+08	& 1.88E+08	& 1.02E+01 \\
140	& 6.93E+11	& 2.96E+08	& 2.66E+08	& 1.31E+00 \\
160	& 7.77E+11	& 4.33E+08	& 3.72E+08	& 1.69E-01 \\
180	& 8.49E+11	& 6.16E+08	& 5.15E+08	& 2.39E-02 \\
200	& 9.15E+11	& 8.42E+08	& 7.06E+08	& 3.10E-03 \\
220	& 9.73E+11	& 1.10E+09	& 9.41E+08	& 4.86E-04 \\
240	& 1.02E+12	& 1.40E+09	& 1.24E+09	& 5.26E-05 \\
260	& 1.07E+12	& 1.72E+09	& 1.60E+09	& 8.30E-06 \\
280	& 1.12E+12	& 2.06E+09	& 2.01E+09	& 1.09E-06 \\
300	& 1.15E+12	& 2.40E+09	& 2.48E+09	& 1.57E-07 \\
\botrule
\end{tabular*}
\end{table}

\newpage


\end{document}